\documentclass[useAMS,usenatbib]{mn2e}

\usepackage{amsmath}
\usepackage{amssymb}
\usepackage[dvipsnames]{xcolor}
\usepackage{graphicx}
\usepackage{graphics}
\usepackage{hyperref}
\hypersetup{
     colorlinks   = true,
     citecolor    = blue
}

\def\apj{{ ApJ}}
\def\apjl{{ApJL}}

\def\aap{{ A\&A}}
\def\aj{{ AJ}}
\def\mnras{{ MNRAS}}

\def\nat {{ Nature}}

\def\procspie{{Proc. SPIE}}

\def\spa{{\mbox{ }}}
\def\mt{\mathrm}
\newcommand{\myemail}{wenbinlu@astro.as.utexas.edu}

\title[EIC emission from jetted TDEs]{External Inverse-Compton Emission
  from Jetted Tidal Disruption Events} 
\author[Lu \& Kumar]
  {Wenbin Lu,$^1$\thanks{\myemail}
  Pawan Kumar,$^1$\thanks{pk@astro.as.utexas.edu}\\
  $^1$Department of Astronomy, University of Texas at Austin, Austin,
TX 78712, USA}
\date{\today}

\pagerange{\pageref{firstpage}--\pageref{lastpage}} \pubyear{2015}
\def\LaTeX{L\kern-.36em\raise.3ex\hbox{a}\kern-.15em
    T\kern-.1667em\lower.7ex\hbox{E}\kern-.125emX}

\begin{document}
\label{firstpage}
\maketitle

\begin{abstract}
The recent discoveries of Sw J1644+57 and Sw J2058+05 show that tidal
disruption events (TDEs) can launch relativistic jets. 
Super-Eddington accretion produces a strong radiation
field of order Eddington luminosity. In a jetted TDE, electrons in the
jet will inverse-Compton scatter the photons from the accretion disk
and wind (external radiation field). Motivated by observations of thermal
optical-UV spectra in Sw J2058+05 and several other TDEs, we assume
the spectrum of the 
external radiation field intercepted by the relativistic jet to be
blackbody. Hot electrons in the jet scatter this thermal radiation and
produce luminosities $10^{45}-10^{48}\rm\  erg \
s^{-1}$ in the X/$\gamma$-ray band.

This model of thermal plus inverse-Compton radiation is applied to Sw
J2058+05. First, we show that the blackbody component in the
optical-UV spectrum most likely has its origin in the
super-Eddington wind from the disk. Then, using the observed 
blackbody component as the external radiation field, we show that the
X-ray luminosity and spectrum are consistent with the inverse-Compton
emission, under the following conditions: (1) the jet Lorentz factor
is $\Gamma\simeq 5-10$; (2) 
electrons in the jet have a powerlaw distribution $dN_e/d\gamma_e
\propto \gamma_e^{-p}$ with $\gamma_{\rm min}\sim1$ and $p=2.4$; (3) the
wind is mildly relativistic (Lorentz factor $\gtrsim 1.5$) and has
isotropic-equivalent mass loss rate $\sim 5\rm\ M_{\odot}\
yr^{-1}$. We describe the implications for jet composition 
and the radius where jet energy is converted to radiation.
\end{abstract}


\section{Introduction}
A tidal disruption event (TDE) occurs when a star passes close
to a massive black hole (BH). \citet{1988Natur.333..523R} described
the basic physics of tidal disruption, where the star's self gravity
causes the exchange of angular momentum. The outer half of the star
gains angular 
momentum and is ejected, and the inner half is left in bound
elliptical orbits. The bound matter circularizes due to shocks and
then accretes onto the BH. If the BH mass $\lesssim 
10^7M_\odot$, the accretion could be highly super-Eddington and is
believed to produce optical-UV to soft X-ray flares with
luminosities $\sim$ Eddington luminosity lasting for weeks
to months \citep[e.g.][]{2009MNRAS.400.2070S,
  2011MNRAS.410..359L}. Recently, many 
TDE candidates were discovered in the optical-UV
\citep[e.g.][]{2012Natur.485..217G, 2014ApJ...780...44C,
  2014MNRAS.445.3263H, 2014ApJ...792...53V, 2014ApJ...793...38A} and
X-rays \citep[e.g.][]{2004ApJ...603L..17K, 2009ApJ...698.1367G,
  2012A&A...541A.106S}. Usually, blackbody radiation at a temperature
of $\sim 10^{4-5}\ K$ and luminosity $\sim 10^{43 - 45} \
erg \ s^{-1}$ is observed.

The recent discoveries of Swift J164449.3+573451 \citep[hereafter Sw
J1644+57, e.g.][]{2011Sci...333..199L, 2011Sci...333..203B,
  2011Natur.476..421B, 2011Natur.476..425Z} and Swift J2058.4+0516
\citep[hereafter Sw J2058+05,][]{2012ApJ...753...77C,
  2015ApJ...805...68P} show that the accretion can launch
relativistic jets which produce bright multiwavelength
emission from radio to X/$\gamma$-ray. Hereafter, we call these events
``jetted TDEs''.
If the X-ray radiation efficiency is 0.1, the isotropic jet
kinetic power reaches $L_{j}\sim  
10^{48} \spa erg \spa s^{-1}$ for 
$\sim 10^6\spa s$ and then decreases roughly as $t^{-5/3}$. 
Modeling of the radio emission from Sw J1644+57 shows that the total
kinetic energy of the disk outflow is $\sim10^{53}\spa erg$
\citep[e.g.][]{2013ApJ...767..152Z, 2013ApJ...770..146B,
  2014ApJ...788...32W, 2015MNRAS.450.2824M}, which means that either 
the jet beaming factor is 
$\sim 0.1$ (half opening angle $\theta_j$ about $30^o$) or the jet being
narrow ($\theta_j\sim 10^o$) but there is another
outflow component carrying $\sim10$ times more energy.

The thermal optical-UV emission could come
from a super-Eddington wind launched from the accretion disk
\citep[e.g.][]{2009MNRAS.400.2070S}. Due to the large optical
depth, photons are advected by electron scattering in the wind. As a
result of adiabatic expansion, the radiation temperature drops to
$\sim 10\spa eV$ at the radius where photons can
escape. \citet{2015ApJ...806..164P} 
proposed that the energy dissipated by shocks from stream-stream
collisions will also produce optical-UV emission
consistent with many TDE candidates. Both models show that the
thermal emission should be ubiquitous in all TDEs and more or less
isotropic. This is supported by comparisons between the TDE rate
selected by optical-UV observations and the rate predicted from
galactic dynamics \cite[e.g.][]{2002AJ....124.1308D,
  2004ApJ...600..149W}. Therefore, in a jetted TDE, we expect a strong
external radiation  field (ERF) surrounding the jet, and electrons in
the jet will inevitably inverse-Compton scatter the ERF.

In this work, we model the ERF simply as a blackbody (motivated by
TDEs found in optical-UV and soft X-ray surveys) and
calculate the luminosity from inverse-Compton scattering of ERF by
electrons in the jet. If the jet has Lorentz factor $\Gamma$ and
electrons have thermal Lorentz factor $\gamma_e$ in the jet comoving 
frame, external photons'
energy will be boosted by a factor of $\sim
\Gamma^2\gamma_e^2$. For typical seed photon energy $10\ eV$ and
bulk Lorentz factor $\Gamma\sim10$, the scattered photons have energy
$\sim\gamma_e^2\ keV$. Therefore, the external 
inverse-Compton (EIC) process produces X/$\gamma$-ray
emission that could be seen by observers with line of sight passing
inside the relativistic jet cone.

One of the biggest puzzles in the two jetted TDEs Sw J1644+57 and Sw
J2058+05 is the radiation mechanism of X-rays \citep[see][for a
thorough discussion of X-ray generation processes in TDE jets]{Crumley2015}
Is it possible that the X-rays are from EIC emission? Thermal
emission from Sw J2058+05 is detected 
in near-IR, optical and UV bands \citep{2012ApJ...753...77C,
  2015ApJ...805...68P}, thanks to the small dust extinction in the
host galaxy ($A_V\lesssim 0.5\spa mag$). Therefore, we use the
observed thermal component as the ERF and test if the X-ray data is
consistent with being produced by the EIC process. 

This work is organized as follows. In section \ref{sec:jet}, we
describe the characteristics of the jet. In
section \ref{sec:model}, we calculate the expected EIC luminosities
from above and below the ERF photosphere. In section
\ref{sec:applications}, we apply the model to Sw J2058+05. We discuss
uncertainties in our model and suggestions for future observations in 
section 5. A summary is given in section 6. Throughout the work, the
convention $X = 10^n X_n$ and CGS units are used. If not specifically
noted, all luminosities and energies are in the isotropic equivalent
sense.

\section{Jet Characteristics}\label{sec:jet}
We assume a baryonic jet with bulk Lorentz factor $\Gamma\gg1$
and half opening angle $\theta_j\ll1$. By ``on-axis observer'', we
mean that the angle between the jet axis and the observer's line of
sight is smaller 
than the relativistic beaming angle $\Gamma^{-1}$. The jet is assumed
to be steady\footnote{Fluctuations of $L_j$
  on a timescale $\sim$ light-crossing time of Schwarzschild
  radius are inevitable and might be the reason for the fast
  variability seen in X-ray. Here, by ``steady'', we mean the averaged
level on timescales $\sim10^6\ s$.} and the (isotropic) kinetic power 
is denoted as $L_j = 10^{48}L_{j,48}\spa erg \spa s^{-1}$. Electron
number density in the lab frame (BH rest frame) is $n_e(R) = L_j/(4\pi
R^2\Gamma m_pc^3)$. Throughout the work, we assume inverse-Compton
scattering by the electrons in the jet has Thomson cross-section $\sigma_T$
(Klein-Nishina suppression is negligible).

 Consider a small radial segment of the jet as a cylinder of
  height $\Delta R$ and radius $\theta_j R$. For external photons
  traveling across the jet in the transverse direction, the
  optical depth is equal to the total number of electrons within this
  cylindrical volume times $\sigma_T$ divided by the area of the side,
  i.e.
\begin{equation}
  \label{eq:10}
  \tau_{j,trvs} = \frac{\pi \theta_j^2 R^2 \Delta R n_e \sigma_T}{2\pi
  \theta_j R \Delta R}  = \frac{1}{2}R\theta_j n_e \sigma_T
= 5.9\times10^{-3} \frac{L_{j,48}\theta_{j,-1}}{R_{15}\Gamma_1}
\end{equation}
We call the radius where $\tau_{j,trvs}=1$ ``self-shielding
radius''
\begin{equation}
  \label{eq:8}
  R_{j,self} = 5.9 \times 10^{12} 
  \frac{L_{j,48}\theta_{j,-1}}{\Gamma_1} \spa cm
\end{equation}
below which external photons cannot penetrate the jet transversely.
For external photons moving in the radial direction towards the origin
(against the jet flow), the optical depth of the jet is
\begin{equation}
  \label{eq:9}
      \tau_{j,r}  = R n_e \sigma_T = 0.117
\frac{L_{j,48}}{R_{15}\Gamma_1}
\end{equation}
The jet becomes transparent in
the radial direction ($\tau_{j,r}=1$) at radius
\begin{equation}
  \label{eq:43}
  R_{j,tr} = 1.17\times 10^{14} \frac{L_{j,48}}{\Gamma_1} \spa  cm
\end{equation}
which is the radius where the jet has largest scattering
cross section. We can see that it is easier
for photons to penetrate the jet in the transverse direction than in
the radial direction, since the jet is narrow. 
Note that $R_{j,tr}$ is different from the ``classical'' jet
photospheric radius \citep[e.g.][]{2000ApJ...530..292M}, which is
based on the optical depth for photons comoving with the jet
\begin{equation}
  \label{eq:7}
  \tau_{j,cmv} \simeq \frac{n_e \sigma_T R}{\Gamma^2} = 1.2\times
  10^{-3} \frac{L_{j,48}}{R_{15}\Gamma_1^3}
\end{equation}
The difference between $\tau_{j,r}$ (Eq. \ref{eq:9}) and
$\tau_{j,cmv}$ (Eq. \ref{eq:7}) is: the former is for photons
moving against the jet flow, so photons can interact with electrons at
all radii from $0$ to $R$; the latter is for
photons moving along the jet flow, so photons can only interact with 
electrons in the local casualty connected
thickness $\sim R/\Gamma^2$.
From Eq.(\ref{eq:7}), we can see that once an external photon is
scattered by a jet electron at radius $\gtrsim 10^{12}\ cm$, it will escape
freely along the jet funnel.

\section{External inverse-Compton Emission}\label{sec:model}
\begin{figure}
  \centering
  \includegraphics[width=0.45\textwidth,
  height=0.2\textheight]{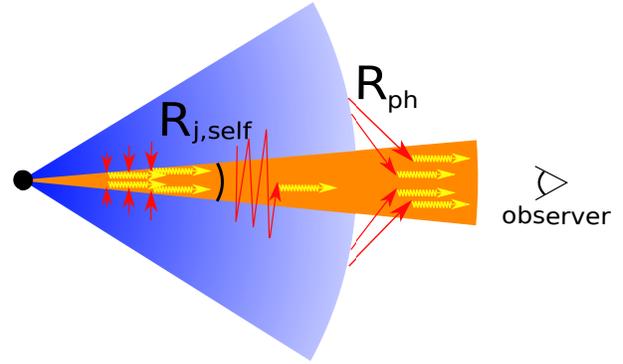}
  \caption{\small Geometry of the external inverse-Compton (EIC)
    process. The material producing the external radiation field (ERF)
    is in \textcolor{blue}{blue} and the jet in
    \textcolor{YellowOrange}{orange}. The observer is on the jet axis.
    The photospheric radius $R_{ph}$ is where the optical depth of the
    ERF emitting material $\tau(R)=1$ (Eq. \ref{eq:94}). The EIC
    scattering could happen above and below $R_{ph}$.
    The self-shielding radius $R_{j,self}$ (Eq. \ref{eq:8}) is where
    the jet becomes transparent in the transverse direction. Usually
    we have $R_{j,self}\ll R_{ph}$, so in the $R_{j,self}<R<R_{ph}$
    region, photons may penetrate the jet transversely multiple times
    before getting scattered.
  }
  \label{fig:photosphere}
\end{figure}

In this section, we construct a simple model for the EIC interaction
between the jet and the ERF, and calculate the EIC luminosities. 
In the jet comoving frame, electrons are assumed to have a 
single Lorentz factor $\gamma_e$. For any distribution of  Lorentz
factors $dN_e/d\gamma_e=N_\gamma(\gamma_e)$, another 
convolution $\int_{\gamma_{min}}^{\gamma_{max}} N_\gamma
d\gamma_e$ is needed (see section \ref{sec:hot_electron}). We assume 
the ERF is emitted from a spherically symmetric photosphere and has a
blackbody spectrum\footnote{Other types of ERF could be produced by
  the accretion disk (multicolor blackbody spectrum), hot corona (disk
  + Comptonization spectrum), shocks (powerlaw spectrum if some
  electrons are accelerated to a powerlaw distribution). They can be
  dealt with by convolving our simple procedure over the ERF
  spectrum.}. 
The photospheric radius of the ERF emitting material $R_{ph}$ is
determined by solving
\begin{equation}
  \label{eq:94}
  \tau(R)= \int_{R}^\infty \kappa \rho dR= 1 
\end{equation}
where $\kappa(R)$ is the total opacity, $\rho(R)$ is the density
profile. If the length-scale of the density gradient $\nabla \rho$ is on the order of
$\sim R$ and $\kappa(R)$ is dominated by electron
scattering $\kappa_s$, the photospheric radius can be estimated by
$\kappa_s \rho(R_{ph})R_{ph}=1$. As shown in
Fig.(\ref{fig:photosphere}), the EIC emission could come from above
and below $R_{ph}$. 

The Rosseland mean absorption opacity (including free-free and
bound-free) is $\kappa_a\sim 10^{25}\rho T^{-3.5}\spa cm^2 \spa g^{-1}$
\citep{1979rpa..book.....R}. The 
density at $R_{ph}$ can be estimated by $\rho\sim  
m_p/(\sigma_TR)\simeq 2\times 10^{-15} R_{15}^{-1} \spa g \spa
cm^{-3}$. Observationally, the temperature at $R_{ph}$ is
$T\gtrsim$ a few $\times 10^4\spa K$. With such a low density and high
temperature, the absorption opacity turns out to be
$\kappa_a\lesssim 10^{-4} \spa 
cm^2 \spa g^{-1}$. Therefore, the opacity is dominated by Thomson
scattering $\kappa_s= 0.34\spa cm^2 \spa g^{-1}$ (assuming
solar metallicity). Note that the radiation at $R_{ph}$ may not be
thermalized, because the ``effective'' absorption optical depth
\citep{1979rpa..book.....R} 
\begin{equation}
  \label{eq:13}
  \tau^*(R) = \int_R^\infty \left[\kappa_a(\kappa_s +
    \kappa_a)\right]^{1/2} \rho dR \sim \rho R \left( \kappa_a \kappa_s
  \right)^{1/2} 
\end{equation}
could be much smaller than 1 at $R_{ph}$. The
``thermalization radius'' $R_{th}$ is defined as where $\tau^*(R_{th})
= 1$ and photons are thermalized only below $R_{th}$.
The ratio $R_{ph}/R_{th}$ (always $>1$) depends on the density
profile. For example, a wind profile
$\rho\propto R^{-2}$ gives $R_{ph}/R_{th} =
(\kappa_s/\kappa_a)^{1/2}\gtrsim 10$. 
Between $R_{th}$ and $R_{ph}$, there's a
purely scattering layer where photons escape via diffusion.  Note that, if the
observed blackbody luminosity and temperature are 
$L_{BB}$ and $T$, the radius determined by $(L_{BB}/4\pi \sigma 
T^4)^{1/2}$ ($\sigma$ being the Stefan-Boltzmann
constant) is usually not the photospheric radius. 

In typical TDEs, the luminosity of the ERF is close to Eddington
luminosity $L_{Edd} \sim 10^{44} (M_{BH} /10^6 M_\odot )\spa erg \spa
s^{-1}$, peaking around optical-UV. With ideal
multiwavelength coverage and small dust 
extinction, the ERF is observable and can be determined by two
parameters: the total luminosity $L_{BB}$ and temperature $T$.
In the following two subsections, we treat $L_{BB}$ and $T$ as knowns.

\subsection{EIC emission from above the photosphere}\label{sec:above}
If the observed blackbody luminosity is $L_{BB}$, the ERF flux at the
photosphere is
\begin{equation}
  \label{eq:19}
  F(R_{ph}) = \frac{L_{BB}}{4\pi R_{ph}^2}
\end{equation}
Since $\tau=1$, the ERF at $R_{ph}$ is not far from being isotropic. At radii
$R>R_{ph}$, the ERF flux drops as 
$R^{-2}$ and photons are moving increasingly parallel with the jet, so 
most of the EIC scatterings happen at radius
$R\sim R_{ph}$ and the (isotropic) EIC luminosity is
\begin{equation}
  \label{eq:20}
  \begin{split}
       L_{EIC}^{(1)} &\simeq \mbox{min}\left(
         \frac{4}{\theta_j^2}, 4\Gamma^2 \right)  \Gamma^2\gamma_e^2
   F(R_{ph}) 2\pi
   R_{ph}^2\theta_j \mbox{min}\left(\tau_{j,trvs}(R_{ph}), 1 \right) \\ 
   &\simeq  \mbox{min}\left(
  1, \theta_j^2\Gamma^2 \right) \Gamma^2\gamma_e^2
\tau_{j,r}(R_{ph}) L_{BB}
  \end{split}
\end{equation}
where $\tau_{j,trvs}$
(Eq. \ref{eq:10}) and $\tau_{j,r}$ (Eq. \ref{eq:9}) are the optical
depth of the jet in the
transverse and radial direction. In the second line of Eq.(\ref{eq:20}),
we have used $\tau_{j,trvs}(R_{ph})<1$, because, for parameter space
relavant to this work, the condition $\tau_{j,trvs}<1$ is
always well satisfied. From Eq.(\ref{eq:20}),
we can see that the EIC process above the photosphere boosts the 
ERF's luminosity by a factor of $\Gamma^2\gamma_e^2
\tau_{j,r}(R_{ph})\sim 10\gamma_e^2$.

\subsection{EIC emission from below the photosphere}\label{sec:below}
 Below the photosphere, the radiation energy in the ERF emitting
  material has a gradient in the direction where the optical
  depth $\tau$ drops, so radiation diffuses outwards at a flux
  \citep{2004rahy.book.....C}
\begin{equation}
  \label{eq:3}
  F_{dif}(R<R_{ph}) \simeq \frac{U(R)c}{3\tau(R)}
\end{equation}
where $U(R)$ is the radiation energy density in the ERF emitting
material at radius $R$ and $c$ is speed of light. As mentioned above,
there is a purely scattering layer between the 
photosphere and thermalization radius. If the ERF emitting material is
  expanding, below the radius where photons are 
  advected by electrons (advection radius $R_{adv}$, see section
  \ref{sec:wind}) or the thermalization radius $R_{th}$, the radiation
  temperature is controlled by adiabatic expansion (assuming radiation
  pressure dominates)
  \begin{equation}
    \label{eq:4}
    U\propto T^4\propto \rho^{4/3} ,\ R<\mathrm{max}(R_{th}, R_{adv})
  \end{equation}
  In the radius range $\mathrm{max}(R_{th}, R_{adv}) < R < R_{ph}$, since
  Comptonization is  
  not efficient enough to change photons' energy, the diffusive flux
  follows the inverse square law from energy conservation
\begin{equation}
  \label{eq:21}
   \frac{Uc}{3\tau}= F(R_{ph}) \left(
    \frac{R}{R_{ph}} \right)^{-2} ,\ \mathrm{max}(R_{th},
  R_{adv}) < R < R_{ph} 
\end{equation}
From Eq.(\ref{eq:94}), (\ref{eq:4}) and (\ref{eq:21}), the radial
distribution of radiation energy $U(R)$ can be solved, once we know
the density profile $\rho(R)$. This is
done in \ref{sec:wind} (Fig. \ref{fig:UR}) under the assumption that
the ERF emitting material is a super-Eddington wind with $\rho\propto
R^{-2}$. A similar discussion is
given in the context of a wind 
from ultra luminous X-ray source M101 X-1 by
\citet{2015MNRAS.447L..60S}. Below, we take $U(R)$ --- the radiation
energy density in the ERF emitting material at polar angle $\theta\gg
\theta_j$ --- as known and consider the energy density in the jet
funnel. 

\begin{figure}
  \centering
  \includegraphics[width=0.45\textwidth,
  height=0.25\textheight]{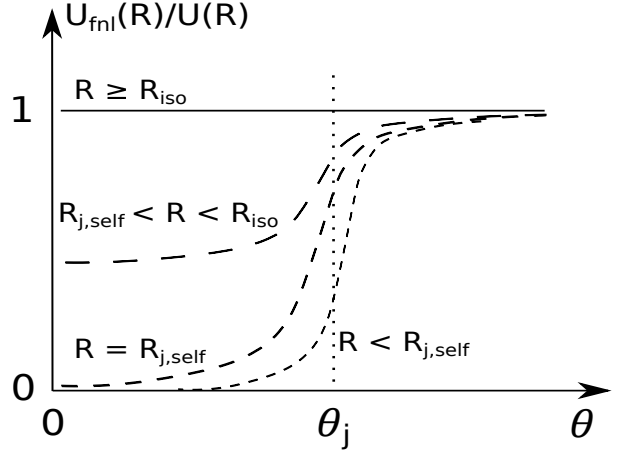}
  \caption{\small A sketch for the radiation energy
    density in the jet funnel $U_{fnl}(R)$ as a function of polar angle
    $\theta$ at different radii R. Above the ``isotropization radius''
    $R_{iso}$ (defined by  Eq. \ref{eq:45}), the removal of photons by the
    jet is balanced by the supplying diffusive flux from the
    surrounding material, so the radiation energy density at the
    funnel $U_{fnl}(R)$ reaches the same as in the surrounding
    material $U(R)$. Below the ``self-shielding radius'' $R_{j,self}$
    (defined by Eq. \ref{eq:8}), the jet is optically thick in the
    transverse direction, so the radiation energy density at the
    funnel center is approximately zero. In between the two
    characteristic radii $R_{j, self}<R<R_{iso}$, the radiation energy
    density in the funnel can be estimated by Eq.(\ref{eq:1}) as
    $U_{fnl}\simeq U/(3\tau \tau_{j,trvs})$.}
\label{fig:Utheta}
\end{figure}

Due to the removal of 
photons by jet scattering, the energy density in the funnel will be
smaller than in the surrounding material far from the funnel. However,
since the jet is narrow, when the optical depth of the jet in the
transverse direction $\tau_{j,trvs}$ is small enough, the radiation
field in the 
funnel will not feel the existence of the jet, i.e. it will isotropize
and reach energy density $U(R)$. We define an ``isotropization
radius'' $R_{iso}$ where the removal of photons by the jet is
balanced by the flux entering the jet funnel $F_{dif}(R)$, i.e.
\begin{equation}
  \label{eq:45}
  \tau_{j,trvs}Uc = \frac{Uc}{3\tau} \mbox{, or }
\tau_{j,trvs}\tau = 1/3
\end{equation}
In the range $R_{j,self}< R < R_{iso}$, the radiation energy density
in the funnel $U_{fnl}(R)$ is smaller than $U(R)$ and is roughly given by
\begin{equation}
  \label{eq:1}
  \tau_{j,trvs}U_{fnl}c \simeq Uc/3\tau
\end{equation}
In the range $R_{iso}<R<R_{ph}$, the radiation energy
density in the funnel equals to $U(R)$. Physically, photons cross the
funnel back and forth in the transverse direction
$1/\tau_{j,trvs}$ times before getting scattered by electrons in the jet, and
when $1/\tau_{j,trvs}\sim \tau$, the radiation field can no longer
distinguish between the funnel and the region far from the funnel and
hence will isotropize. Fig. (\ref{fig:Utheta}) roughly shows the changing
of radiation energy density $U(R)$ with polar angle $\theta$ at
different radii $R$.

The order of $R_{th}$, $R_{adv}$ and
$R_{iso}$ depends on the density profile $\rho(R)$, jet Lorentz factor
$\Gamma$ and jet kinetic power $L_{j}$. In the case of a wind density
profile $\rho\propto R^{-2}$ in the TDE context, we typically have
$R_{th}\lesssim R_{adv}\sim R_{iso}$ (see section
\ref{sec:wind}). The
EIC luminosity below the photosphere is mostly produced at radius
$R\sim R_{iso}$ and we have
\begin{equation}
  \label{eq:17}
    \begin{split}
       L_{EIC}^{(2)} &\simeq \mathrm{min}\left(
         \frac{4}{\theta_j^2}, 4\Gamma^2 \right)  \Gamma^2\gamma_e^2
   F_{dif}(R_{iso}) 2\pi R_{iso}^2\theta_j \\
   &= \mathrm{min}\left(
     1, \theta_j^2\Gamma^2 \right)
   \frac{2\Gamma^2\gamma_e^2}{\theta_j} L_{BB}\  \mathrm{min} \left[ 1,
     \left( \frac{R_{iso}}{R_{adv}} \right)^{1/3} \right]
  \end{split}
\end{equation}
where we have normalized the diffusive flux at $R_{iso}$ to the total
luminosity $L_{BB}$ by $F_{dif}(R_{iso})4\pi
R_{iso}^2 =L_{BB}\ \mathrm{min} \left[ 1,
     \left( R_{iso}/R_{adv} \right)^{1/3} \right]$.
The EIC scattered photons' peak energy is
$\Gamma^2\gamma_e^22.82kT\mbox{ max}[1, (R_{iso}/R_{adv})^{-2/3}]$. 
Eq.(\ref{eq:17}) means that the EIC process below the 
photosphere boosts the ERF's luminosity by a factor of
$2\Gamma^2\gamma_e^2/\theta_j \sim (100-1000) \gamma_e^2$.

\subsection{Corrections for mildly relativistic wind}
If the ERF comes from a super-Eddington wind launched from the disk,
the wind velocity $\beta_w = v_w/c$ could be mildly relativistic. In this
subsection, we show that relativistic effects make the EIC scattered
photons' energy and EIC luminosities
(Eq. \ref{eq:20} and \ref{eq:17}) smaller. Depending on $\beta_w$, the
corrections could be significant. Quantities in the wind
comoving frame are denoted by a prime ($^\prime$) and those in the lab
frame are unprimed.

If the wind Lorentz factor is $\Gamma_w =
(1-\beta_w^2)^{-1/2}$, the relative Lorentz
factor between the jet and wind is $\Gamma_{rel} = \Gamma
\Gamma_w(1-\beta \beta_w)\simeq \Gamma\Gamma_w(1-\beta_w)$. For
example, if $\Gamma = 10$, $\beta_w = 0.3$ $(0.8)$ gives $\Gamma_{rel} =
7.3$ $(3.4)$. After EIC scattering, external photons' energy is only
boosted by a factor of $\Gamma_{rel}^2\gamma_e^2$, which could be much
smaller than $\Gamma^2\gamma_e^2$.

If the observed blackbody luminosity and temperature are $L_{BB}$ and
$T$, the radiation energy density at the wind photosphere
$R_{ph}$ in the wind comoving frame is
\begin{equation}
  \label{eq:5}
  U^\prime (R_{ph}) = \frac{L_{BB}}{4\pi R_{ph}^2 c\Gamma_w^2}
\end{equation}
The wind photospheric radius is different from the non-relativistic
case of Eq.(\ref{eq:94}) by a factor $\Gamma_w^2$ and is given by
\begin{equation}
  \label{eq:2}
  \tau_w(R) = \kappa_s \rho_{w} (R)R/\Gamma_w^2 = 1
\end{equation}
where the rest mass density $\rho_w$ is related to the (rest) mass
loss rate by mass conservation
\begin{equation}
  \label{eq:54}
    4\pi R^2 \rho_w(R) v_w = \dot{M}_w
\end{equation}
Therefore, the EIC luminosity from above the
photosphere is 
\begin{equation}
  \label{eq:6}
  \begin{split}
      L_{EIC}^{(1)} &\simeq \mt{min}(1, \theta_j^2\Gamma^2)
  \Gamma_{rel}^2\gamma_e^2 \tau_{j,r}(R_{ph}) 4\pi R_{ph}^2
  U^{\prime}(R_{ph}) c \\
  & = \mt{min}(1, \theta_j^2\Gamma^2)
  \Gamma_{rel}^2\gamma_e^2 \tau_{j,r}(R_{ph}) L_{BB}/\Gamma_w^2
  \end{split}
\end{equation}

The EIC luminosity from below the photosphere is mostly produced at
the isotropization radius $R_{iso}$ and can be estimated as
 \begin{equation}
   \label{eq:12}
   \begin{split}
        L_{EIC}^{(2)} &\simeq \mt{min}\left(
         \frac{4}{\theta_j^2}, 4\Gamma^2 \right)  \Gamma_{rel}^2\gamma_e^2 
   F_{dif}^\prime(R_{iso}) 2\pi R_{iso}^2\theta_j \\
   & = \mt{min}\left(
     1, \theta_j^2 \Gamma^2 \right)
   \frac{2\Gamma_{rel}^2\gamma_e^2}{\theta_j}
   \frac{L_{BB}}{f(\Gamma_w)} 
   \mathrm{min} \left[ 1,
     \left( \frac{R_{iso}}{R_{adv}} \right)^{1/3} \right]
   \end{split}
 \end{equation}
Here, the normalization from the diffusive flux
$F_{dif}^\prime(R_{iso}) = U^\prime (R_{iso})c/3\tau_w(R_{iso})$ to
$L_{BB}$ is different from the 
non-relativistic case used in Eq. (\ref{eq:17}) by a factor of 
\begin{equation}
  \label{eq:14}
  f(\Gamma_w) = \Gamma_w^2 (1 - \beta_w/3)(1+\beta_w)^3
\end{equation}
which will be derived in section \ref{sec:wind}. The
EIC scattered photons' peak energy is
\begin{equation}
  \label{eq:39}
  h\nu_{EIC} = 
  \begin{cases}
    & \Gamma_{rel}^2\gamma_e^22.82kT \mbox{, above the photosphere} \\
    &\Gamma_{rel}^2\gamma_e^22.82kT \mathrm{max}\left[1,
      \left(\frac{R_{iso}}{R_{adv}}\right)^{-2/3}\right]  \mbox{,
      below\dots}
  \end{cases}
\end{equation}

\section{Applications to Sw J2058+05}\label{sec:applications}
Similar to the more widely studied event Sw J1644+57, Sw
J2058+05 has a rich set of data, in terms of multiwavelength
(radio, near-IR, optical, UV, X-ray, $\gamma$-ray) and time coverage (a
few to $\sim200$ days, in the host galaxy rest frame). In this
section, we use the 
data published by \citet{2012ApJ...753...77C, 2015ApJ...805...68P} and
test if the X-rays from Sw J2058+05 are consistent with the EIC emission
from the jet. We focus on Sw J2058+05 because it suffers from a small
amount of host galaxy dust extinction and reddening ($A_V\lesssim
0.5 \spa mag$, while Sw J1644+57 has $A_V\sim 5\spa mag$). All
quantities (time, frequencies and luminosities) are 
measured in the host galaxy rest frame at redshift $z = 1.185$
\citep{2012ApJ...753...77C}. 

The X-ray lightcurve and spectrum of Sw J2058+05 are similar to Sw
J1644+57. The main X-ray properties are as follows:
(1) The isotropic luminosity stays
$\gtrsim 10^{47}\spa erg\spa s^{-1}$ for $\sim 20\spa d$ and then
decline as $\sim t^{-5/3}$ until a sudden drop (by a factor $>160$)
at $\sim 200\spa d$. (2) Rapid variability ($\lesssim 500\spa s$) is
detected before the drop off, suggesting the X-ray emitting region is
at radius $R\sim 10^{15}  
(\delta t/500\spa s) (\Gamma/10)^2 \spa cm$. (3) The spectra could be
fit by an absorbed powerlaw, with early time
($25-86\spa d$, from {\it Swift}/XRT) spectral index $\alpha \sim
0.5$ ($\nu L_\nu\propto \nu^{\alpha}$) and late time ($100 - 200\spa d$)
$\alpha \sim 0.2-0.3$. We note that the early 
time index $\alpha \sim  0.5$ comes from combining\footnote{Similar to
  Sw J1644+57, Sw 2058+05 could have different spectral indexes at different
  flux levels \citep{2012MNRAS.422.1625S}. However,
  single  {\it Swift}/XRT observations do not have enough statistics
  to constrain the 
  spectral parameters in Sw 2058+05.} all the XRT PC-mode
data within $25-86\spa d$, and hence should be taken with
caution. We use $\alpha =0.3$ as a typical spectral index in the
following. 

\begin{figure}
  \centering
  \includegraphics[width=0.5\textwidth,
  height=0.25\textheight]{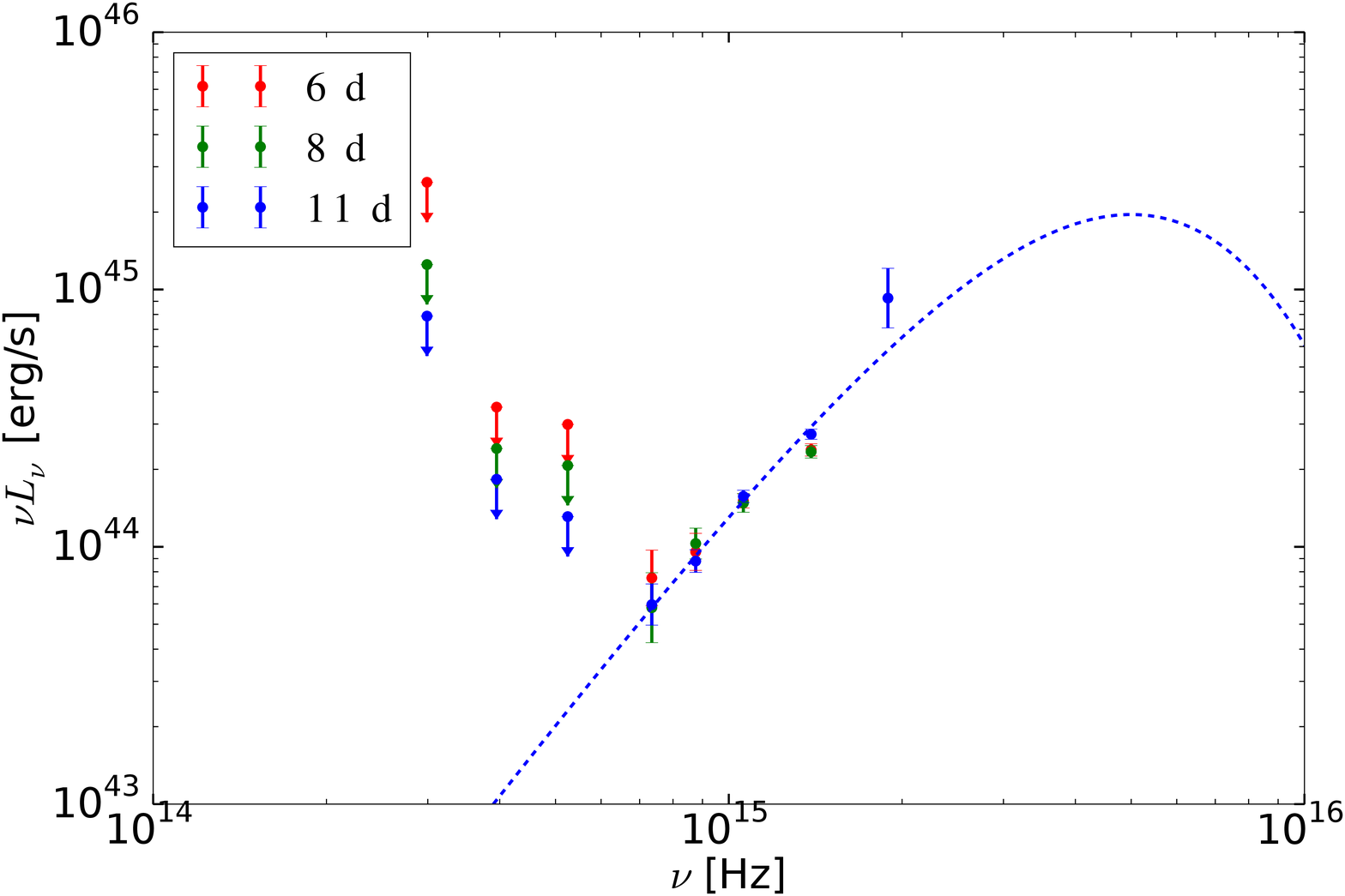}
  \caption{\small The early time optical-UV spectra of Sw
    J2058+05. They are consistent with 
    blackbody. The \textcolor{blue}{blue} dashed line represents
  a blackbody spectrum of temperature $T= 6\times 10^4\spa K$,
  normalized by $\nu L_\nu (10^{15} \spa Hz) = 1.3\times 10^{44} \spa
  erg \spa s^{-1}$. Since the data points only cover the
  Rayleigh-Jeans tail, the adopted temperature is a lower
  limit of the true blackbody temperature. All quantities (time,
  frequencies and luminosities) are 
    measured in the host galaxy rest frame. }
  \label{fig:spectraEarly}
\end{figure}

The reported optical-UV magnitudes are not
corrected for dust extinction. We correct the reddening from the Milky
Way (in the direction of this event), using $E(B-V) = 0.095\spa mag$
\citep[][and refs therein]{2012ApJ...753...77C}. The extinction $A_b$
in any band $b$ is calculated by using the tabulated $A_b/E(B-V)$
value (at $R_V=3.1$) from \citet{2011ApJ...737..103S}. The host
galaxy is at redshift $z= 1.185$, so the luminosity distance is $2.54\times
  10^{28} \spa cm$, if a standard $\Lambda$CDM cosmology is assumed
  with $H_0 = 71 \spa km \spa s^{-1} \spa Mpc^{-1}$, $\Omega_m =
  0.27$, and $\Omega_{\Lambda} = 0.73$. We refer to the time of
  discovery as 00:00:00 on MJD = 55698, following 
\citet{2012ApJ...753...77C}. The rest-frame time is estimated as
$(\mbox{time} - 55698)/(1+z)$. We use the effective wavelengths
$\lambda_{eff}$ of different filters and the rest-frame
frequencies are calculated by $\nu = (1+z)c/\lambda_{eff}$.

\begin{figure}
  \centering
  \includegraphics[width=0.5\textwidth,
  height=0.25\textheight]{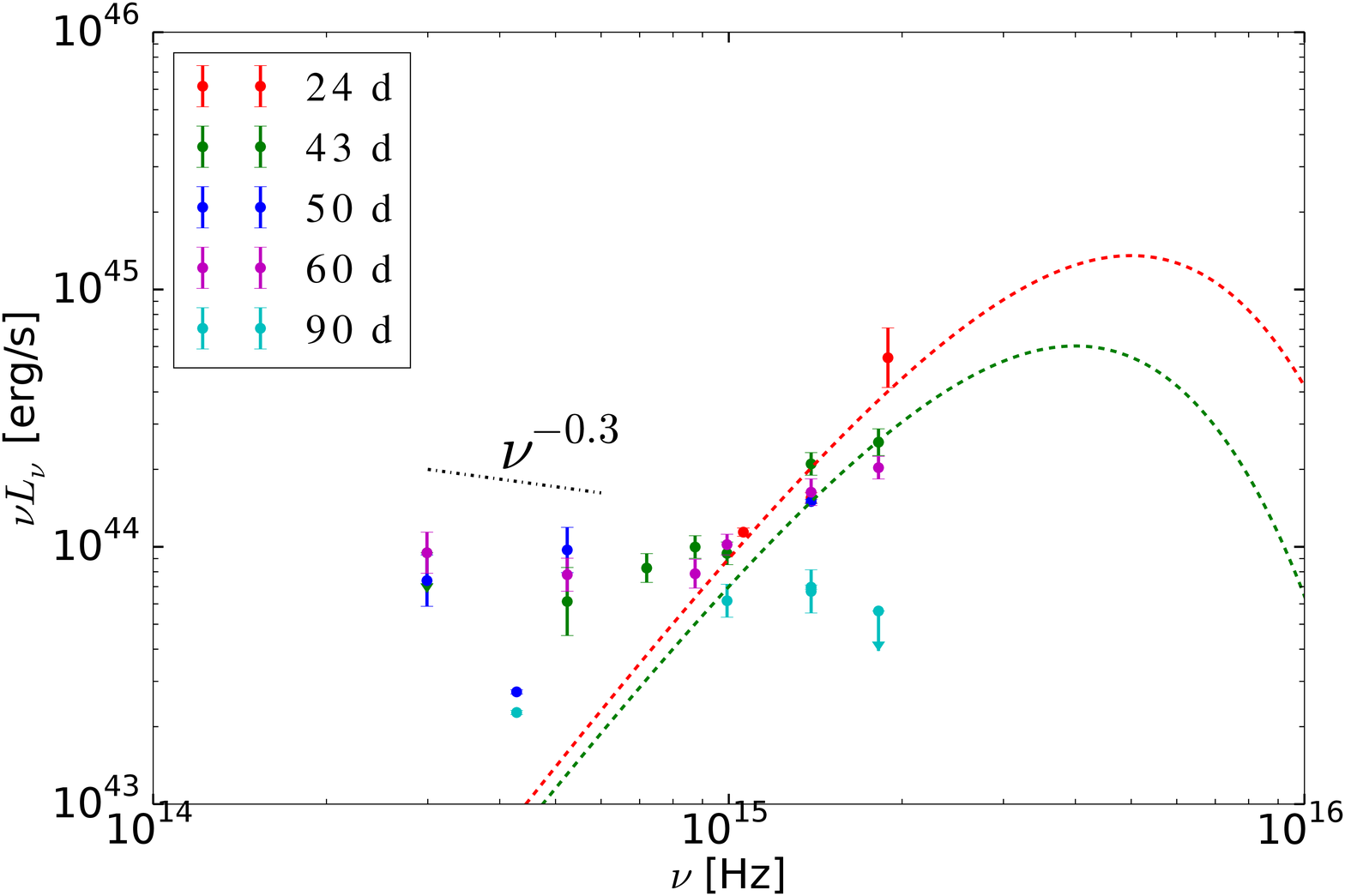}
  \caption{\small The optical-UV spectra of Sw J2058+05 from
    $24 \spa d$ to $90 \spa d$. Apart from the blackbody spectrum, a 
    powerlaw component shows up on the low frequency end. The dashed
    lines are blackbodies with parameters 
    \{$T=6\times10^4\spa K$, $\nu L_\nu (10^{15} \spa
    Hz) = 9\times 10^{43} \spa erg \spa s^{-1}$\} (\textcolor{red}{red}) and
    \{$T=5\times10^4\spa K$, $\nu L_\nu (10^{15} \spa 
    Hz) = 7\times 10^{43} \spa erg \spa s^{-1}$\}
    (\textcolor{OliveGreen}{green}). Since the 
    data points only cover the Rayleigh-Jeans tail, the adopted
    temperatures are lower limits. The black dotted line is a
    representative powerlaw 
    $\nu L_\nu\propto \nu^{-0.3}$ (not a fit to the data). Considering
    the large 
    errorbars and uncertainties from host galaxy reddening, a
    powerlaw of index $\alpha \sim -0.3$ to
    $\sim 0.5$ ($\nu L_\nu \propto \nu^\alpha$) could fit the
    data. 
    We note that the two data points at
    $4.3\times10^{14}\spa 
    Hz$ (from {\it
      HST/F160W}, see Table 2 of \citet{2015ApJ...805...68P}) seem
    not consistent with the powerlaw, which could be due to the under
    estimation of errorbars. However, the blackbody component is not
    affected. All quantities (time, frequencies and luminosities) are 
    measured in the host galaxy rest frame.} 
  \label{fig:spectraMid}
\end{figure}

\begin{figure}
  \centering
  \includegraphics[width=0.5\textwidth,
  height=0.25\textheight]{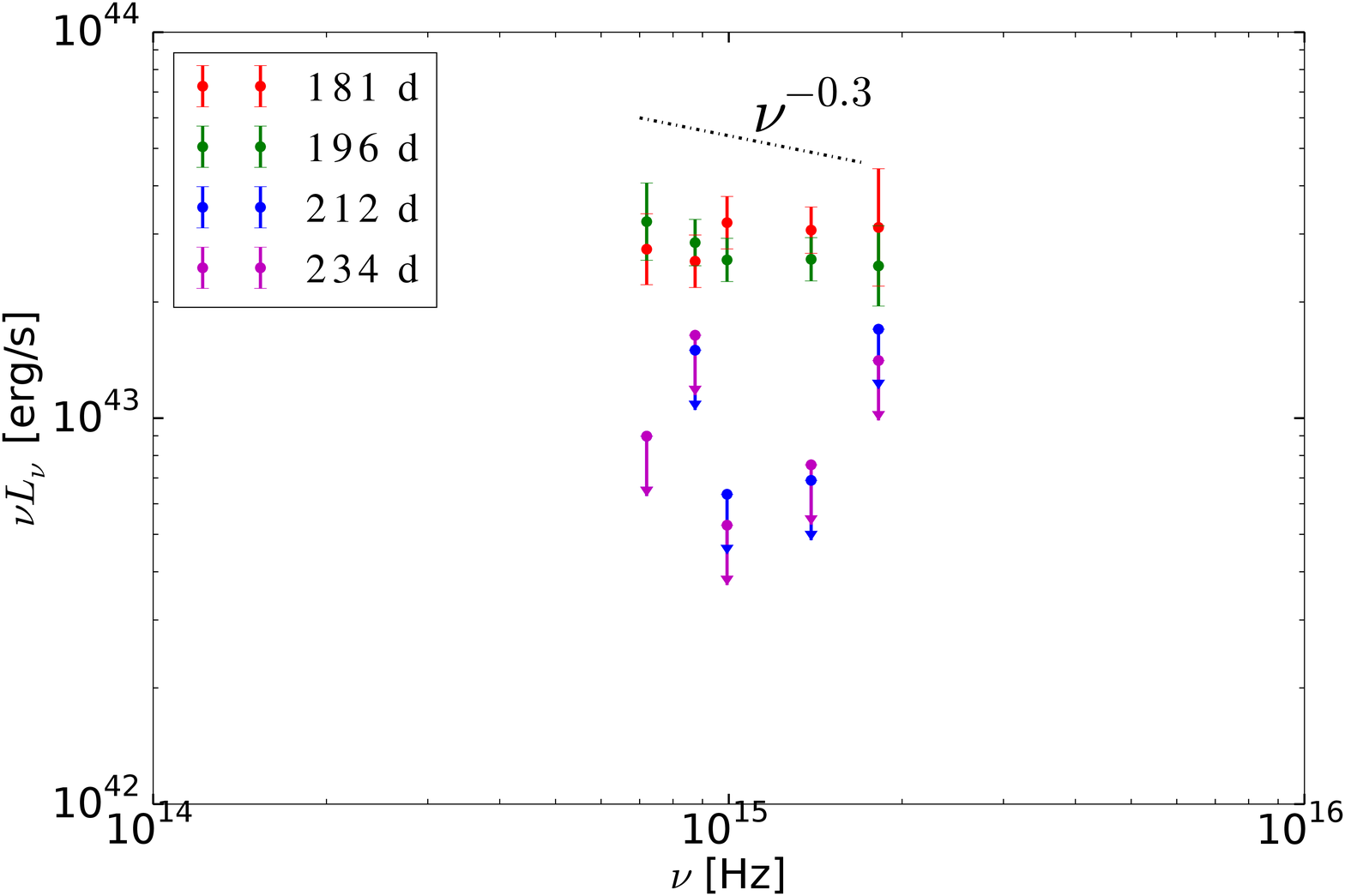}
  \caption{\small The optical-UV spectra of Sw J2058+05 from
    $181\spa d$ to $234\spa 
    d$. There is no visible blackbody component, so we can get an
    upper limit $\nu
    L_\nu(2\times10^{15}\spa Hz) < 2\times 10^{43}\ erg\ s^{-1}$.
    Considering the large
    errorbars and uncertainties from host galaxy reddening, a
    powerlaw of index $\alpha \sim -0.3$ to
    $\sim 0.5$ ($\nu L_\nu \propto \nu^\alpha$) could fit the
    data. The black dotted line is a 
    representative powerlaw 
    $\nu L_\nu\propto \nu^{-0.3}$ (not a fit to the data). All
    quantities (time, frequencies and luminosities) are  
    measured in the host galaxy rest frame.}
  \label{fig:spectraLate}
\end{figure}

The optical-UV spectra at different time are shown in
Fig.(\ref{fig:spectraEarly}, \ref{fig:spectraMid},
\ref{fig:spectraLate}). The spectrum is 
purely a blackbody at early time $t\simeq 6-11\spa d$, 
then a powerlaw component shows up on the low 
frequency end at $t\simeq 20-60\spa d$, and when $t\gtrsim 100\spa d$,
the powerlaw component
dominates and the blackbody component becomes invisible. For our
purpose, we focus on the blackbody component 
hereafter (see section \ref{sec:discussion} for a discussion about the
powerlaw component).

A blackbody spectrum can be described by two parameters, the bolometric
luminosity $L_{BB}$ and the temperature $T$, as follows
\begin{equation}
  \label{eq:22}
  L_\nu = \frac{15 h L_{BB}}{\pi^4 kT}
  \frac{(h\nu/kT)^3}{exp(h\nu/kT)-1} 
\end{equation}
where $h$ is the Planck constant and $k$ is the Boltzmann constant.
Unfortunately, optical-UV observations only cover the Rayleigh-Jeans tail,
which is insufficient to fully describe a blackbody spectrum. From
Fig.(\ref{fig:spectraEarly}) and (\ref{fig:spectraMid}), we can get
two pieces of information: (i)
a lower limit on the temperature
\begin{equation}
  \label{eq:25}
     T \geq 6 \xi \times 10^4 \spa K
\end{equation}
where $\xi = 1$, $1$, $0.8$ when $t = (6-11)$, $24$,
$(43-60)\spa d$, respectively; (ii) the normalization
\begin{equation}
  \label{eq:26}
     \nu L_\nu (10^{15}\spa Hz) = \chi \times 10^{44}\spa
     erg \spa s^{-1}
\end{equation}
where $\chi = 1.3$, $0.9$, $0.7$ when $t = (6-11)$, $24$,
$(43-60)\spa d$, respectively.
Making use of Eq.(\ref{eq:22}), we can rewrite Eq.(\ref{eq:26}) as
\begin{equation}
  \label{eq:36}
  \begin{split}
      L_{BB} = &6.5\times 10^{44}\chi
      \left(\frac{T}{4.8\times10^4\spa K} 
  \right)^4 \\
  &\left[exp(4.8\times10^4\spa K/T) - 1
\right] \spa erg\spa s^{-1}
  \end{split}
\end{equation}
where $T$ is the blackbody temperature (constrained by
Eq. \ref{eq:25}). Hereafter, we use the approximation
  $x^4[exp(1/x) - 1] \simeq 1.4x^{3}$, which is accurate to $\leq 20\%$
  when $x\in (1,5)$. Eq.(\ref{eq:25}) and (\ref{eq:36}) are all
the information we can get from the observed spectra.

In Fig.(\ref{fig:spectraLate}), there's no visible blackbody
component from $181\spa d$ to $234\spa d$, so we get an
upper limit $\nu L_\nu(2\times10^{15}\spa Hz) < 2\times
10^{43}erg/s$. The jet might have been turned off at this time,
because the X-ray sharp drop occurs at $t\sim 200\spa d$.

We note that the host galaxy may contribute a small amount of 
reddening\footnote{\citet{2015ApJ...805...68P} fit the {\it
    XMM-Newton} X-ray 
  ($0.3-10\spa keV$) spectra with a single powerlaw and obtain an
  absorbing column density. After subtracting the Galactic column
  density $0.088\times 10^{22}\spa cm^{-2}$
  \citep{2005A&A...440..775K}, we get $N_H(\mbox{host})\simeq 1\pm
  1.5\times 10^{21}\spa cm^{-2}$.} similar to the Milky Way, which
will make the spectra slightly steeper, but the conclusions
on the blackbody component (Eq. \ref{eq:25} and \ref{eq:36})  are only
mildly affected. These uncertainties could be taken into account by
the two dimensionless parameters $\xi$ and $\chi$.

In the following two subsections, we first show that the blackbody
component can be produced by a super-Eddington
wind. Next, we use the observed blackbody
component as the ERF and test if the X-ray lightcurve and spectrum are
consistent with the EIC emission from above or below the
photosphere. Constraints on 
the jet parameters from the two cases are summarized in
Table (\ref{tab:summary}). Note that, since the EIC model in section
\ref{sec:model} is under the assumption of the jet being
ultra-relativistic ($\Gamma\gg 1$), if the constraints lead to
$\Gamma\lesssim 2$, the model is inconsistent with the data.

\subsection{Wind Model}\label{sec:wind}
The high X-ray luminosity of Sw J2058+05 implies that the accretion stays
super-Eddington for a few months. Super-Eddington
disks are known to be accompanied by strong winds. For 
instance, \citet{2007MNRAS.377.1187P} show that strong winds combined
with the X-rays from the disk around super-Eddington accreting
stellar-mass BHs are in good agreement of the observational data from 
ultra luminous X-ray sources. The super-Eddington wind could be
launched by radiation pressure
\cite[e.g.][]{1973A&A....24..337S}. Rencent
radiation-magnetohydrodynamic (rMHD) simulations by 
\citet[][2D]{2011ApJ...736....2O} and
\citet[][3D]{2014ApJ...796..106J} show that the kinetic power of 
(continuum) radiation driven wind can be much higher than
$L_{Edd}$. However, the 3D general relativistic rMHD
simulations by \citet{2014MNRAS.441.3177M} show that the kinetic power
of wind from super-Eddington disks around rapidly spinning BHs remains
at the order of $L_{Edd}$. \citet{2014MNRAS.438.3024L} proposed
that the strength of line driven winds sharply rises when the local
temperature of the accretion disks around supper massive BHs reaches
$\sim 5\times 10^4\ K$.  It is also likely that magnetic fields (MFs) play an
important role in the wind launching process, since angular momentum
is removed from an accretion disk through MFs. For example,
\citet{1982MNRAS.199..883B} proposed that the wind could be driven
centrifugally along open MF lines.

Up to now, a systematic study of the role of MFs and (line-
and continuum-) opacity is still
lacking and the detailed wind launching 
physics is still not well understood. In the context of TDEs, the fact
that the fall-back material is very weakly bound is very different
from the initial conditions used in the aforementioned numerical
simulations. Since the fall-back material evolves nearly
adiabatically, the energy released from the accretion of a fraction of
the material on bound orbit could push the rest outwards as a wind.

Hereafter, we use upper case $R$ to denote the true radii (in $cm$)
and lower case $r = R/R_S$ for the dimensionless radii normalized by
the Schwarzschild radius $R_S = 
3\times 10^{11} m_6\spa cm$. Also, the true
accretion, outflowing (subscript ``w''), and fallback 
(subscript ``fb'') rates (in $M_\odot \spa yr^{-1}$) are denoted as upper case
$\dot{M}$ and the dimensionless rates are normalized by the Eddington
accretion rate as $\dot{m} = \dot{M}/\dot{M}_{Edd}$. The Eddington
accretion rate is defined as $\dot{M}_{Edd} = 
10L_{Edd}/c^2$,  and $L_{Edd} =1.5\times10^{44} m_6\spa erg \spa
s^{-1}$, where $m_6=M/10^6M_\odot$ is BH mass in $10^6M_\odot$ and
we have assumed solar metallicity with Thomson scattering opacity
$\kappa_s = 0.34\ cm^2\ g^{-1}$.

For a star with mass $M_* = m_* M_\odot$ and radius $R_* = r_*
R_\odot$, the (dimensionless) tidal disruption radius is
\begin{equation}
  \label{eq:52}
    r_T = \frac{R_*}{R_S} \left(    \frac{M}{M_*} \right)^{1/3}
  \simeq 23.3 m_6^{-2/3} m_*^{-1/3} r_*
\end{equation}
The star's original orbit has pericenter distance $r_p<r_T$. When the
star passes $r_T$ for the first time, the tidal force from 
the BH causes a spread of specific orbital energy across the star
\citep{2013MNRAS.435.1809S} 
\begin{equation}
  \label{eq:95}
  \Delta \epsilon = \frac{GM}{R_T}\frac{R_*}{R_T}
\end{equation}
Bound materials have specific orbital energies $-\Delta \epsilon <
\epsilon <0$ and the corresponding Keplerian orbital periods $P$ are
given by
\begin{equation}
  \label{eq:96}
  \epsilon = -\frac{1}{2} \left( \frac{2\pi GM}{P}
\right)^{2/3}
\end{equation}
Therefore, if circularization is efficient enough (within a few
orbital periods), the fall back rate is
\begin{equation}
  \label{eq:97}
  \dot{M}_{fb} = \frac{dM_*}{dP} = \frac{dM_*}{d\epsilon}
  \frac{d\epsilon}{dP} = \frac{( 2\pi GM)^{2/3}}{3}
  \frac{dM_*}{d\epsilon} P^{-5/3}
\end{equation}
which means that a flat distribution of mass per orbital energy gives
the mass fall-back rate $\dot{M}_{fb}\propto
(t/t_o)^{-5/3}$. The leading edge of the
fall-back material has the shortest period
\begin{equation}
  \label{eq:98}
  t_o = 41 m_6^{1/2} m_*^{-1} r_*^{3/2} \spa d
\end{equation}
Therefore, the normalized fall-back rate profile is
\begin{equation}
  \label{eq:99}
  \dot{m}_{fb} = 1.12\times 10^2 m_6^{-3/2} m_*^2 r_*^{-3/2}
  (t/t_o)^{-5/3}
\end{equation}
Following \citet{2009MNRAS.400.2070S}, we assume a fraction
$f_{out}\in(0.1,1)$ of the fall-back gas is gone with the wind, and
hence the wind mass loss 
rate is $\dot{m}_{w}\sim10-100$ at early time ($\lesssim 20 \spa d$) and
$\dot{m}_{w}\propto t^{-5/3}$ later on (if $f_{out}$ stays
constant). Note that, in the absence of the wind,
the jet might be draged to a halt due to the IC scattering of
radiation from the disk as follows. From the conservation of angular
momentum, the disk size is $2r_p\sim r_T\sim 10^{13}\ cm$, which is
larger than the 
self-shielding radius $R_{j.self}$ (Eq. \ref{eq:8}). Therefore, at
$R=10^{13}\ cm$, disk photons penetrate the jet funnel in the
transverse direction and hence the inverse-Compton power of each
electron in the jet is $P_{IC} \simeq L_{disk}/(4\pi
R^2)\sigma_T\Gamma^2\gamma_e^2$. The ratio of EIC drag timescale,
$t_{IC} = \Gamma m_p c^2/P_{IC}$ (assuming electrons and protons are
coupled), and the dynamical timescale, $t_{dy} = R/2c$, is
\begin{equation}
  \label{eq:37}
  \frac{t_{EIC}}{t_{dy}} = 1.7 \frac{R_{13}}{L_{disk,45}
    \Gamma_1 \gamma_e^2}
\end{equation}
As we show in this paper, an optically thick mildly relativistic wind
alleviates this IC drag problem and links the observed optical-UV to
the X-ray emission in a self-consistent way.

We assume that the wind is launched from radius $r_o=R_o/R_S$ at a
speed $\beta_w = v_w/c$. Due to inadequate understanding of the wind
launching physics, the radius $r_o$ is uncertain and hence taken as a
free parameter in this work. The rMHD simulations mentioned at the
beginning of this subsection show that $r_o\sim$ a few.

\begin{figure}
  \centering
  \includegraphics[width=0.45\textwidth,
  height=0.25\textheight]{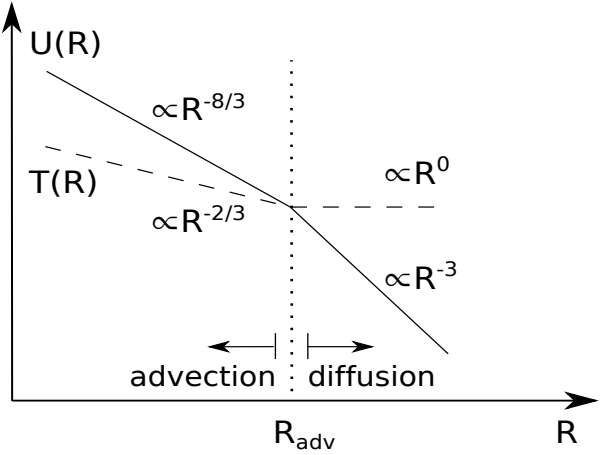}
  \caption{\small A sketch for the evolution of radiation energy
    density $U(R)$ and temperature $T(R)$ with radius $R$. Below the
    ``advection radius'' $R_{adv}$ (defined by Eq. \ref{eq:59}),
    the energy density is controlled by adiabatic expansion. Above
    $R_{adv}$, the energy density decreases with $R$ because of
    diffusion. Since Comptonization is not efficient enough to change 
    photons' energy, the radiation temperature stays constant at
    $R>R_{adv}$.}
  \label{fig:UR}
\end{figure}

At the wind launching radius $r_o$, we assume that 
radiation energy and kinetic energy are in equipartition:
\begin{equation}
  \label{eq:53}
    4\pi R_o^2\Gamma_w^2 U^\prime(R_o) v_w = (\Gamma_w -
    1)\dot{M}_wc^2
\end{equation}
The radiation temperature at the base of the wind $T_o^\prime$ is
related to the radiation energy density by $U^\prime(r_o) 
= a(T_o^\prime)^4$ ($a$ being the radiation density constant), so from
Eq.(\ref{eq:53}), we have 
\begin{equation}
  \label{eq:55}
  T_o^\prime \simeq 4.9\times 10^{6} \left(\frac{\Gamma_w -
      1}{\Gamma_w\beta_w} \right)^{1/4}
  r_o^{-1/2} \dot{m}_{w,2}^{1/4} m_6^{-1/4} \spa K
\end{equation}
Combining Eq. (\ref{eq:2}) and (\ref{eq:54}), we obtain the
photospheric radius of the wind
\begin{equation}
  \label{eq:58}
    r_{ph} \simeq \frac{5.0\times10^2
      \dot{m}_{w,2}}{\Gamma_w^2\beta_w}
\end{equation}
Below $r_{ph}$, photons escape by diffusion or advection, and the
radius where diffusion time equals to the dynamical time
(i.e. $\tau_w=c/v_w$) is called the ``advection radius''
\begin{equation}
  \label{eq:59}
    r_{adv} \simeq \frac{5.0\times10^2 \dot{m}_{w,2}}{\Gamma_w^2}
\end{equation}
At smaller radii $r<r_{adv}$, the wind evolves adiabatically, so the
radiation pressure, which dominates over gas pressure ($nkT$),
decreases with density as $P=a T^4/3 \propto \rho^{4/3}$. Under the
assumption of a steady wind  
with constant velocity and spherical symmetry, the density profile is
$\rho\propto r^{-2}$, so the radiation temperature (in the comoving
frame) evolves as 
\begin{equation}
  \label{eq:15}
  T^\prime (r) = T_o^\prime (r/r_o)^{-2/3} \mt{\ if\ r_o < r < r_{adv}}
\end{equation}
Here, at a temperature $\gtrsim 10^5\ K$, the thermalization radius
$r_{th}$ (defined by $\tau^*(r_{th})=1$ according to Eq. \ref{eq:13})
is related to the photospheric radius by
$r_{ph}/r_{th}=(\kappa_s/\kappa_a)^{-1/2}\gtrsim 10$. Since
$r_{ph}/r_{adv}=c/v_w\lesssim 10$, we usually have $r_{th}\lesssim
r_{adv}$. In the range $r_{adv}<r<r_{ph}$, photons only
interact with baryons by electron scattering (or Comptonization),
which is not efficient enough to change photons' energy
significantly. Therefore, the radiation temperature stays constant as
\begin{equation}
  \label{eq:16}
  T^\prime(r) = T^\prime_{adv} = T_o^\prime (r_{adv}/r_o)^{-2/3} \mt{\
  if\ r_{adv}<r<r_{ph}}
\end{equation}
Combining Eq.(\ref{eq:55}), (\ref{eq:59}) and (\ref{eq:16}), we find
the radiation temperature at the advection radius $T_{adv}^\prime$ (in
the wind comoving frame). The blackbody temperature to be
observed\footnote{Strictly speaking, the spectrum integrated over the
  whole photosphere is not Plankian, because the temperature is a
  function of latitude angle $\theta$ (see Eq. \ref{eq:28} below). The
  blackbody approximation makes the equations explicitly solvable and
  hence greatly simplifies the model. We have verified that the error
  in the integrated spectrum resulting from the blackbody
  approximation is less than $40\%$, if $\Gamma_w\lesssim 2$.} is  
$T_w \simeq \Gamma_w T_{adv}^\prime$ and is given by
\begin{equation}
  \label{eq:23}
      T_{w} \simeq 7.8\times 10^{4} \Gamma_w^{11/6} \left(\frac{\Gamma_w -
      1}{\beta_w} \right)^{1/4} r_{o}^{1/6} m_{w,2}^{-5/12} m_6^{-1/4} \mt{\ K}
\end{equation}
In the range $r_{adv}<r<r_{ph}$, photons escape by diffusion
and the diffusive flux follows the inverse square law $F_{dif}^\prime
= U^\prime c/3\tau_w \propto r^{-2}$ (since radiation energy is
conserved), so we have 
\begin{equation}
  \label{eq:27}
  U^\prime(r) = U^\prime(r_{adv}) (r/r_{adv})^{-3} \mt{\ if\ r_{adv}<r<r_{ph}}
\end{equation}
The evolution of radiation energy density and
temperature with radius in the wind model is shown in
Fig.(\ref{fig:UR}). 

Next, we Doppler-boost the radiation field from the wind
comoving frame to the lab frame to calculate the luminosity seen by
the observer. The specific intensity at $r_{ph}$ in the wind comoving
frame is
\begin{equation}
  \label{eq:24}
  \begin{split}
      I^\prime_{\nu^\prime}(r_{ph}) &= I^\prime_{\nu^\prime}(r_{adv})
  \left(\frac{r_{ph}}{r_{adv}} \right)^{-3} \\
  & = \beta_w^3 \frac{2h (\nu^\prime)^3}{c^2}
  \frac{1}{\mt{e}^{h\nu^\prime /kT_{adv}^\prime} - 1}
  \end{split}
\end{equation}
After Lorentz transformation $I_\nu = I^\prime_{\nu^\prime}
(\nu/\nu^\prime)^3$, the specific intensity in the lab frame is still
a blackbody and the only difference is that the temperature is a
function of the emission latitude angle $\theta$, i.e.
\begin{equation}
  \label{eq:28}
  I_\nu(r_{ph}, \theta) = \beta_w^3 \frac{2h \nu^3}{c^2}
  \frac{1}{\mt{e}^{h\nu /k\tilde{T}(\theta)} - 1}
\end{equation}
where $\tilde{T}(\theta) = T_{adv}^\prime/[\Gamma_w(1 - \beta_w \cos
\theta)]$. Note that the difference between relativistic and
non-relativistic solutions is the latitude dependence of
$\tilde{T}(\theta)$, and the flux ratio is a function of wind Lorentz
factor
\begin{equation}
  \label{eq:30}
  \begin{split}
      f(\Gamma_w) &= \frac{\int_0^1 \mu d\mu
        \int d\nu I_\nu(\mu) 
      }{\int_0^1 \mu d\mu \int d\nu^\prime I^\prime_{\nu^\prime}}
      = \frac{\int_0^1 \mu [\Gamma_w(1 - \beta_w \mu)]^{-4} d\mu
}{\int_0^1 \mu d\mu}\\
    & = \Gamma_w^2(1 - \beta_w/3)(1+\beta_w)^3
  \end{split}
\end{equation}
where $\mu = \cos \theta$ has been used. Note that
$f(\Gamma_w)\rightarrow 16\Gamma_w^2/3$ in the ultra-relativistic
limit and $f(\Gamma_w)\rightarrow 1$ in the non-relativistic
limit. The isotropic equivalent luminosity for an observer at infinity is
\begin{equation}
  \label{eq:29}
  \begin{split}
          L_{w} &= 4\pi \int_0^1 2\pi
  R_{ph}^2 \int I_\nu(R_{ph}, \mu) d\nu \mu d\mu\\
  & = \pi R_{adv}^2 a (T_{adv}^\prime)^4 \beta_w c f(\Gamma_w)\\
 &\simeq 5.9\times 10^{44}\Gamma_w^{1/3}(\Gamma_w - 1) f(\Gamma_w)
 r_{o}^{2/3} \dot{m}_{w,2}^{1/3} m_6 \ erg\ s^{-1}
  \end{split}
\end{equation}                                              
from which, we can see that  the wind
luminosity can mildly exceed the Eddington luminosity (when $\dot{m}_w
\gg1$).
Putting the optical-UV constraints from Sw J2058+05 (Eq. \ref{eq:25} and
\ref{eq:36}) into the wind model (Eq. \ref{eq:23} and \ref{eq:29}), we
find 
\begin{equation}
  \label{eq:64}
  \begin{split}
        &m_6 \dot{m}_{w,2}^{19/21} = 2.9
        \frac{\Gamma_w^{38/21}}{[f(\Gamma_w)]^{4/7} (\Gamma_w-1)^{1/7} 
      \beta_w^{3/7}} \chi^{4/7} r_{o}^{-2/21} \\
    &\dot{m}_{w,2}  \leq 0.97 \frac{[f(\Gamma_w)]^{3/4} \Gamma_w^{29/4} 
      (\Gamma_w-1)^{3/2}}{\beta_w^{3/4}}  \xi^{-21/4} \chi^{-3/4} r_{o} 
  \end{split}
\end{equation}
where $\xi = 1$, $1$, $0.8$ and $\chi = 1.3$, $0.9$, $0.7$ when $t =
(6-11)$, $24$, $(43-60)\spa d$, respectively.
We note that, due to the strong dependence on the temperature
(through $\xi$) and wind velocity $\beta_w$, the upper limit of mass loss
rate $\dot{m}_w$ has large uncertainties and so does the lower
limit of BH mass $m$. However, the product
$m_6\dot{m}_{w,2}\simeq m_6 \dot{m}_{w,2}^{19/21}$ only
depends on $\beta_w$, decreasing from $\sim4$ to 2 when $\beta_w\in
(0.3, 0.99)$. Therefore, the true wind mass loss rate can be estimated by
\begin{equation}
  \label{eq:65}
  \dot{M}_{w} = 2.6 m_6\dot{m}_{w,2} \spa M_\odot\spa yr^{-1}\simeq
  5 \frac{m_6\dot{m}_{w}}{200} \spa  M_\odot\spa yr^{-1}
\end{equation}
Note that the derived mass loss rate $\dot{M}_{w}$ is in the isotropic
equivalent sense. The wind is expected to be somewhat beamed along the
jet axis (towards the observer), so Eq.(\ref{eq:65}) is consistent with a
typical TDE and the optical-UV blackbody component is consistent with
being produced by a super-Eddington wind. Note that the advection
radius only depends on the product $m_6\dot{m}_{w,2}$ and is hence not
affected by the uncertainties on the temperature:
\begin{equation}
  \label{eq:56}
  R_{adv} = 3.0\times10^{14} \frac{1}{\Gamma_w^2}\frac{m_6
    \dot{m}_{w}}{200}  \ cm 
\end{equation}
And the photospheric radius $R_{ph}$ is a factor $1/\beta_w$ larger.

In section \ref{sec:below}, we defined the ``isotropization radius''
$R_{iso}$ by balancing the radiation flux entering the jet funnel
through the interface with the wind and the flux removed due to EIC
scattering. In the relativistic case, $R_{iso}$ is given by
\begin{equation}
  \label{eq:11}
  \tau_{j,trvs}^\prime (R)\tau_w(R) = 1/3
\end{equation}
where $\tau_{j,trvs}^\prime (R) = R\theta_jn_e\sigma_T/2\Gamma_w$ is
the transverse optical depth of the jet in the wind comoving frame and
$\tau_w(R) = \kappa_s\rho_w(R)R/\Gamma_w^2$ is the optical depth of
the wind. Below $R_{iso}$, all the diffusive flux $F_{dif}^{\prime}$ 
entering the jet funnel is scattered by the jet and contributes
to the EIC luminosity. At radii $R_{iso}<R<R_{ph}$, the removal of
radiation by EIC scattering is not efficient enough, so the radiation
energy density in the funnel reaches the same as in the wind region
far away from the funnel. Solving Eq.(\ref{eq:11}), we get
\begin{equation}
  \label{eq:67}
  R_{iso} \simeq 7.3\times10^{13} 
\frac{1}{\Gamma_w^{3/2} \beta_w^{1/2}}
\left(
    \frac{L_{j,48}\theta_{j,-1}}{\Gamma_1} 
    \right)^{1/2} \left(
      \frac{m_6\dot{m}_{w}}{200}  \right)^{1/2} \spa cm
\end{equation}

\subsection{EIC Model}\label{sec:hot_electron}
At radii $R<R_{adv}$, the ERF temperature evolves as
$T\propto R^{-2/3}$, so the EIC emission is expected to have a
powerlaw spectrum
\begin{equation}
  \label{eq:73}
  \frac{dL_{EIC}}{d\nu} \propto \frac{dL_{EIC}}{dT} =
  \frac{dL_{EIC}}{dR} \frac{dR}{dT}\propto R \propto
  T^{-3/2}
\end{equation}
from which we get $\nu L_\nu\propto \nu^{-1/2}$. This is too soft
compared to the observed X-ray powerlaw $\nu L_\nu\propto
\nu^{0.3}$. Below, we consider the electrons in the jet having a
powerlaw distribution function
\begin{equation}
  \label{eq:31}
  \frac{dN_e}{d\gamma_e} 
  \begin{cases}
    \propto \gamma_e^{-p} &\mbox{ if
    }\gamma_{min}<\gamma_e<\gamma_{max}\\ 
    =0 &\mbox{ otherwise}
  \end{cases}
\end{equation}
The ERF is assumed to have a blackbody spectrum at temperature
$T$ and bolometric luminosity $L_{BB}$, so the scattered photons'
spectrum at frequency $\nu\gg \Gamma_{rel}^2\gamma_{min}^2 kT/h$ will
be $\nu L_\nu \propto \nu^{(3-p)/2}$. Therefore, the observed X-ray
spectrum $\nu L_\nu \propto \nu^{0.3}$ can be reproduced by an
electron index of $p=2.4$. 

Another requirement is that the $\nu^{0.3}$ powerlaw extends wider
than the $0.3(1+z)-10(1+z)\spa keV$ window. We define
two (electrons') Lorentz factors $\gamma_1$ and $\gamma_2$
corresponding to the scattered photons' energies
\begin{equation}
  \label{eq:40}
  \begin{cases}
    &h\nu_{EIC}(\gamma_e = \gamma_1)= 0.3 (1+z)\spa keV \\
    &h\nu_{EIC}(\gamma_e = \gamma_2)= 10 (1+z)\spa keV\\
    & \gamma_{min} \leq \gamma_1 < \gamma_2 \leq \gamma_{max} 
  \end{cases}
\end{equation}
where $2.82kT$ is the blackbody peak energy and $h\nu_{EIC}$ is given
by Eq.(\ref{eq:39}). We focus on the
XRT band, because the possible extension in the BAT band (up to
$\sim 150(1+z)\spa keV$) could be explained by simply extending
$\gamma_{max}$ to larger values (but $\gamma_{max}$  
is finite so that the EIC luminosity doesn't diverge).

As pointed out in section \ref{sec:model}, the EIC emission
could come from above or below the photosphere. The only difference
is that the EIC luminosity from below the photosphere is larger
by a factor of $2\Gamma_w^2/[\theta_j
\tau_{j,r}(R_{ph})f(\Gamma_w)]\sim 10$ (see Eq. \ref{eq:6} 
and \ref{eq:12}). In the following two subsections, we consider the
two possibilities and try to match the expected EIC luminosities in
the $0.3(1+z)-10(1+z)\spa keV$ window with the observation $L_{X} = 
10^{47}L_{X,47}\spa erg \spa s^{-1}$.

\subsubsection{EIC emission from above the
  photosphere}\label{sec:hot_above} 
In this subsection, we consider the EIC emission
from above the photosphere. We convolve Eq.(\ref{eq:6}), where
electrons are assumed to have a single Lorentz factor $\gamma_e$, with
the Lorentz factor distribution described by Eq.(\ref{eq:31}). Then we
match the EIC luminosity in the $0.3(1+z)-10(1+z)\spa 
keV$ window with observations
\begin{equation}
  \label{eq:86}
    \begin{split}
     L_{X} &\simeq
      \mbox{min}(1,\theta_j^2 \Gamma^2) 
      \Gamma_{rel}^2\gamma_{min}^2\tau_{j,r}(R_{ph}) L_{BB} \\
      &\cdot \frac{p-1}{3-p} \left(\frac {\gamma_1}{\gamma_{min}}
        \right)^{3-p} 
        \left[\left( \frac{\gamma_2}{\gamma_1}
          \right)^{3-p} - 1\right] \\
        &\simeq 10^{47}L_{X,47} \spa erg \spa s^{-1}
  \end{split}
\end{equation}
Combining the X-ray constraints (Eq. \ref{eq:40} and \ref{eq:86})
with optical-UV constraints (Eq. \ref{eq:25} and \ref{eq:36}), we get
\begin{equation}
  \label{eq:84}
  \begin{cases}
    & T/K\simeq
    \frac{1.0\times10^5 \Gamma_w^{0.74}}{\left(
        \Gamma_{rel}\gamma_{min} 
      \right)^{0.52}} 
    \left[ \frac{L_{X,47}}{\mathrm{min}(1, \theta_j\Gamma^2) \chi
      \tau_{j,r}(R_{ph})} \right]^{0.37} \\
      &T/K \leq 2.7\times10^6\left(
          \Gamma_{rel}\gamma_{min} \right)^{-2}\\ 
      &T/K \geq 6\xi \times10^4
  \end{cases}
\end{equation}
Then, we eliminate the parameter $T$ and put the constraints on the
Lorentz factors 
\begin{equation}
  \label{eq:85}
  \begin{cases}
    & \Gamma \gamma_{min}\left[
        \mbox{min}(1,\theta_j^2 
        \Gamma^2) \right]^{-0.25}  
      \leq \frac{9.1}{\Gamma_w^{1.5}(1-\beta_w)} \left(
        \frac{L_{X,47}}{\chi \tau_{j,r}(R_{ph})} 
        \right)^{-0.25}\\
     &\Gamma \gamma_{min}
     \left[ \mbox{min}(1,\theta_j^2 \Gamma^2)
      \right]^{0.71}\leq \frac{2.7\Gamma_w^{0.43}}{1-\beta_w} \left(
        \frac{L_{X,47}}{\chi \tau_{j,r}(R_{ph})} 
        \right)^{0.71} \xi^{-1.9}
  \end{cases}
\end{equation}
The uncertainty lies on the parameter $\tau_{j,r}(R_{ph})$ (the
optical depth of the jet in the radial direction at the ERF's
photosphere $R_{ph}$). Combining Eq.(\ref{eq:9}) and (\ref{eq:58}),
we have
\begin{equation}
  \label{eq:33}
  \tau_{j,r}(R_{ph}) = 0.39\Gamma_w^2\beta_w \frac{L_{j,48}}{\Gamma_1}
  \left( \frac{m_6 \dot{m}_w}{200} \right)^{-1}
\end{equation}

\subsubsection{EIC emission from below the
  photosphere}\label{sec:hot_below}
In this subsection, we consider the EIC emission from below the
photosphere. Similar to the treatment
in section \ref{sec:hot_above}, we match the EIC luminosity in the
$0.3(1+z)-10(1+z)\spa keV$ window with observations
\begin{equation}
  \label{eq:83}
  \begin{split}
      L_X&\simeq \mathrm{min}(1,\theta_j^2\Gamma^2)
      \frac{2\Gamma_{rel}^2\gamma_1^2}{\theta_j}
      \frac{L_{BB}}{f(\Gamma_w)}
      \mathrm{min} \left[ 1,
        \left(\frac{R_{iso}}{R_{adv}}\right)^{1/3}\right] \\
      &\cdot \frac{p-1}{3-p} \left(\frac {\gamma_1}{\gamma_{min}}
        \right)^{3-p} 
        \left[\left( \frac{\gamma_2}{\gamma_1}
          \right)^{3-p} - 1\right]\\
        &\simeq 10^{47}L_{X,47} \spa erg \spa s^{-1}
  \end{split}
\end{equation}
Combining the X-ray constraints (Eq. \ref{eq:40} and \ref{eq:83})
with optical-UV constraints (Eq. \ref{eq:25} and \ref{eq:36}), we get
\begin{equation}
  \label{eq:91}
  \begin{cases}
    &T/K \simeq \frac{7.8\times10^4}{(\Gamma_{rel}\gamma_{min})^{0.52}}
    \left[ 
      \frac{L_{X,47}\theta_j f(\Gamma_w)}{\chi \mathrm{min}(1,\theta_j^2
        \Gamma^2) \mathrm{min} \left[ 1,
        \left(R_{iso}/R_{adv} \right)^{0.53}\right] } \right]^{0.37}\\
      &T/K \leq 2.7\times10^6 (\Gamma_{rel} \gamma_{min})^{-2}
      \mathrm{min} \left[ 1,
        \left(\frac{R_{iso}}{R_{adv}}\right)^{2/3}\right] \\ 
      &T/K \geq 6\xi \times10^4 
  \end{cases}
\end{equation}
We eliminate
the parameter $T$ and put the constraints on the Lorentz factors
\begin{equation}
  \label{eq:92}
  \begin{cases}
    &\Gamma \gamma_{min} \left[\frac{\theta_j}{\mathrm{min}(1,\theta_j^2
        \Gamma^2)}  
      \right]^{0.25}\leq \frac{10.9 \left( \chi/L_{X,47}
        \right)^{0.25}  \mathrm{min} \left[ 1,
        \left(R_{iso}/R_{adv}\right)^{0.58}\right]  }{\Gamma_w(1-\beta_w)
        [f(\Gamma_w)]^{0.25}}  \\
     &\Gamma \gamma_{min}
     \left[\frac{\theta_j}{\mathrm{min}(1,\theta_j^2 \Gamma^2)}
      \right]^{-0.71}\leq \frac{1.68[f(\Gamma_w)]^{0.71}
\left(L_{X,47}/\chi \right)^{0.71} }{\Gamma_w
        (1-\beta_w) \xi^{1.9} \mathrm{min} \left[ 1,
        \left(R_{iso}/R_{adv}\right)^{0.38}\right]}
  \end{cases}
\end{equation}
The ratio of the isotropization radius $R_{iso}$ to the advection
radius $R_{adv}$ can be calculated from Eq.(\ref{eq:56}) and
(\ref{eq:67})
\begin{equation}
  \label{eq:68}
  \frac{R_{iso}}{R_{adv}}\simeq 0.24 \left(
    \frac{\Gamma_w}{\beta_w} \right)^{1/2} \left(
    \frac{L_{j,48}\theta_{j,-1}}{\Gamma_1} 
    \right)^{1/2} 
   \left( \frac{m_6\dot{m}_{w}}{200}  \right)^{-1/2} 
\end{equation}
which means $R_{iso}\lesssim R_{adv}$.

\subsubsection{Results}\label{sec:results}
\begin{table*}
 \begin{minipage}{0.95\textwidth}
   \centering
  \caption{Summary of the constraints on the jet parameters from
    the EIC models above and
    below the photosphere. The three observables $\xi$ (blackbody
    temperature, Eq. \ref{eq:25}),
    $\chi$ (normalization, Eq. \ref{eq:26}), $L_{X,47}$ (X-ray
    luminosity in the $0.3-10\ keV$ window) are obtained by fitting
    the data by hand and have uncertainties 
    $\lesssim 30\%$, so the constraints are accurate to
    within a factor of $\sim2$. Due to various uncertainties such as host
    galaxy dust extinction and X-ray absorbed powerlaw fitting, it's
    hard to achieve a better accuracy anyway. We consider four
    different wind velocities $\beta_w = v_w/c =0.1,\ 0.3,\ 0.6,\ 0.8$
    ($\Gamma_w =  
    (1-\beta_w^2)^{-1/2}$ is the wind Lorentz factor). We can see that,
    for a slow wind with $\beta_w\lesssim 0.6$, the EIC model from
    above the photosphere is consistent with 
    observations but that from below the photosphere is inconsistent
    (marked in \textcolor{red}{red}). The
    physical reason is that the latter over-produces the X-ray
    luminosities at all or some of the epochs. On the other hand, for a
    fast wind with $\Gamma_w\gtrsim 1.5$, the EIC models from both above
    and below the photosphere are consistent with
    observation, with reasonable jet parameters $\Gamma\simeq
    5-10$, $\gamma_{min} \sim 1$ and $p = 2.4$.}
  \label{tab:summary}
  \begin{tabular}{@{}ccccccccccccc@{}}
  \hline\hline
   $\beta_w$  & \multicolumn{3}{c}{$0.1\ (\Gamma_w-1 =$ 5.0e-3)}
    & \multicolumn{3}{c}{$0.3\ (\Gamma_w-1 =$ 4.8e-2)}
                & \multicolumn{3}{c}{$0.6\ (\Gamma_w =1.25)$}
    &\multicolumn{3}{c}{$0.8\ (\Gamma_w=1.67)$}\\
\hline
   $t/d$  & $6-11$ & 24 & $43-60$
    & $6-11$ & 24 & $43-60$
                          & $6-11$ & 24 & $43-60$
          & $6-11$ & 24 & $43-60$\\
 \hline
$\xi$ & 1& 1 & 0.8 & 1& 1 & 0.8 & 1& 1 & 0.8 & 1& 1 & 0.8\\
$\chi$ & 1.3  & 0.9  & 0.7 & 1.3  & 0.9  & 0.7 & 1.3  & 0.9  & 0.7 & 1.3  & 0.9  & 0.7 \\
$L_{X,47}$ & 4.8  & 0.8 & 8.5e-2
                          & 4.8  & 0.8 & 8.5e-2
                           & 4.8  & 0.8 & 8.5e-2
          & 4.8  & 0.8 & 8.5e-2 \\
\hline
\multicolumn{13}{c}{EIC model above the photosphere,
    from Eq.(\ref{eq:85})}\\
\hline
$\Gamma^{5/4} \gamma_{min} \leq$
& 5.0 & 4.8 & 4.7 
& 10.4 & 10.0 & 9.8
& 21.7 & 20.9 & 20.4
& 37.1 & 35.7 & 34.7\\
\hline
$\Gamma^{2/7} \gamma_{min} \leq$
& 22.8 & 25.5 & 42.3
& 6.2 & 7.0 & 11.6
& 3.4 & 3.8 & 6.3
& 3.5 & 3.9 & 6.5\\
\hline
\multicolumn{13}{c}{EIC model below the photosphere,
    from Eq.(\ref{eq:92})}\\
\hline
$\Gamma^{4/3} \gamma_{min} \leq$
& 14.0 & 14.7 & 15.8
& 15.1 & 15.8 & 17.0
& 18.3 & 19.2 & 20.6
& 22.9 & 23.9 & 25.7\\
\hline
$\Gamma^{4/3} \gamma_{min} \leq$
& \textcolor{red}{2.0} &  \textcolor{red}{1.0} &  \textcolor{red}{0.56}
& \textcolor{red}{3.7} & \textcolor{red}{1.8} & \textcolor{red}{1.0}
& 9.4 & 4.6 & \textcolor{red}{2.5}
& 23.8 & 11.6 & 6.4\\
\hline\hline
\end{tabular}
\end{minipage}
\end{table*}

Eq.(\ref{eq:85}) and (\ref{eq:92}) are the general constraints
on the EIC emission models from above and below the
photosphere. However, too many unknown parameters are involved,
including $\Gamma$, $\gamma_{min}$, $\theta_j$, $\Gamma_w$,
$\tau_{j,r}(R_{ph})$, and $R_{iso}/R_{adv}$. To express 
the constraints in a more clear way, we relax some
generalities and make two additional assumptions 
\begin{equation}
  \label{eq:34}
  \begin{cases}
    & L_{j,48} = L_{X,47} \\
    & \theta_j = \Gamma^{-1}
  \end{cases}
\end{equation}
We have to be cautious not to over-interpret the results, because the
two assumptions in Eq.(\ref{eq:34}) are not derived from first
principles. The wind mass loss rate in Eq.(\ref{eq:64}) can be safely
simplified by dropping the $r_o^{-2/21}$ term and ignoring the
difference between $\dot{m}_{w,2}^{19/21}$ and $\dot{m}_{w, 2}$.

At three different epochs ($t = 6-11, 24$ and $43-60$ d), we put the
observables $\xi$ (blackbody temperature, Eq. \ref{eq:25}), $\chi$ 
(normalization, Eq. \ref{eq:26}), $L_{X,47}$ (X-ray luminosity in the
$0.3-10\ keV$ window) into Eq.(\ref{eq:85}) and (\ref{eq:92}), and obtain
the constraints on the two Lorentz factors $\Gamma$ and
$\gamma_{min}$, as summarized in Table \ref{tab:summary}. From the
variability time $\Gamma\simeq 6 (R/10^{15}cm)^{1/2}
(\delta t/500s)^{-1/2}$ and radio beaming \citep[$\Gamma\geq
2.1$][]{2012ApJ...753...77C} arguments, the jet must be relativistic. If the
product $\Gamma\gamma_{min}$ is restricted to be $\lesssim2$, the model
is not consistent with observations. We note that the unphysical result
$\Gamma\gamma_{min}<1$ appears because we assume
the jet is ultra-relativistic ($\Gamma\gg 1$) and it simply means
the EIC process over-produces the X-ray luminosity.

We find: (1) for a slow wind with $\beta_w\lesssim 0.6$, the EIC model from
above the photosphere is consistent with observations but that from
below the photosphere is inconsistent. The physical reason is that the
latter over-produces the X-ray luminosities at all or some of the
epochs. (2) For a fast wind with $\Gamma_w\gtrsim 1.5$, the
EIC models from both above and below the photosphere are consistent
with observation, with reasonable jet parameters $\Gamma\simeq 5-10$,
$\gamma_{min} \sim 1$ and $p = 2.4$.

\section{Discussion}\label{sec:discussion}
In this section, we discuss some potential issues for the EIC
scenario proposed in this work.

(i) The X-ray spectral evolution is not considered in the simple model
described in this work. For
Sw J2058+05, late time ($100-200\spa d$) {\it XMM-Newton} observations
don't show significant change in the spectral slope and 
{\it Swift}/XRT observations don't have enough statistics to constrain
the spectral slope. However, for Sw J1644+57, significant spectral
changes are found when the flux fluctuates on short ($\sim
1\spa d$)  timescale and as the mean flux level evolves on long ($\sim
100\spa d$) timescale \citep{2012MNRAS.422.1625S}. Specifically, the spectrum
is softer $\nu L_\nu\propto \nu^{\sim 0.3}$ at early epochs
($<50\spa d$) and harder $\nu L_\nu\propto \nu^{\sim 0.6}$ later
on. In the EIC scenario, this hardening could be explained
by the following two possibilities: (1) when the accretion rate is
smaller at later time, the ERF comes from smaller radii and has a harder
spectrum; (2) the electrons' powerlaw becomes harder at later
time. Another issue is whether  
the X-ray spectrum is always a single powerlaw in the $0.3(1+z) -
10(1+z) \spa keV$ window. For example, if we repeat the same
procedure in 
section \ref{sec:hot_electron} in a narrower window, e.g. $1(1+z) -
10(1+z) \spa keV$, the constraints will be weaker. {\it Swift}/XRT
observations have too low statistics to pin down this uncertainty, but
future wide field-of-view X-ray telescopes will find more jetted TDEs
\citep{2015ApJ...803...36D}, and with simultaneous optical-UV
coverage, the EIC scenario could be tested to a higher accuracy.

(ii) Another issue is whether the electrons can maintain a powerlaw
distribution. The magnetization of the jet $\sigma$ is defined as the
ratio of magnetic energy over baryons' kinetic energy. The strength of
magnetic field in the jet comoving frame is
\begin{equation}
  \label{eq:102}
  B^\prime
  \simeq 8.2 \times 10^2 \frac{[L_{j,48}\mbox{min}(1,
    \sigma)]^{1/2}}{\Gamma_1 R_{15}} \  G
\end{equation}
The synchrotron cooling time can be estimated as $t_{syn}^\prime =
\gamma_e m_ec^2/P_{syn}^\prime$, where $P_{syn}^\prime$ is the
synchrotron power. Therefore, the ratio of synchrotron cooling time
over dynamical time is
\begin{equation}
  \label{eq:103}
  \frac{t_{syn}^\prime}{t_{dy}^\prime} \simeq 0.70 \frac{1}{\gamma_e}
  \frac{\Gamma_1^3 R_{15}}{L_{j,48} \mbox{min}(1, \sigma)}
\end{equation}
Apart from synchrotron cooling, electrons also suffer from
inverse-Compton (IC) cooling by scattering X-ray photons, 
which have a  
comoving energy density $U_x^\prime = L_X/(4\pi R^2c\Gamma^2)$. The IC
cooling time can be estimated as $t_{IC}^\prime = \gamma_e
m_ec^2/P_{IC}^\prime$, 
where $P_{IC}^\prime$ is the IC power. Therefore, the ratio of IC
cooling time over dynamical time is
\begin{equation}
  \label{eq:104}
  \frac{t_{IC}^\prime}{t_{dy}^\prime} \simeq 7.0 \frac{1}{\gamma_e}
  \frac{\Gamma_1^3 R_{15}}{L_{X,47}}
\end{equation}
At $t \simeq 6-11\ d$, we have $L_{X,47}\simeq 5$, so nearly all electrons
are in the fast cooling regime (due to either
synchrotron or IC cooling). Here, we have used the X-ray radiation
field as a conservative estimate of the IC cooling time and the
optical-UV photons cause even faster IC cooling. We note 
that, in the EIC model since $\gamma_{min} \sim 1$, electrons only
share a very small fraction of the total jet energy at radius
$R\sim10^{14} - 10^{15}\ cm$. Magnetic reconnection
or some non-Coulomb interactions between protons and electrons
may keep reheating the electrons and maintain the powerlaw
distribution.

(iii) Better blackbody temperature measurements or constraints are
crucial. The constraints from the two models (Eq. \ref{eq:85} and
\ref{eq:92}) are both sensitive to the blackbody temperature (through
the parameter $\xi$). For Sw J1644+57, high dust extinction prevents us
from measuring the temperature accurately. However, up to now,
the (small number) statistics show that one out of the two jetted
TDEs has low dust extinction, so  better temperature measurements in
the future might be promising. For Sw J2058+05, due to various
uncertainties such as photometric
measurements, host galaxy reddening, X-ray powerlaw fitting and
crudeness of our model, the constraints on $\Gamma$, $\gamma_{min}$ in
Table \ref{tab:summary} are accurate to a factor of 
$\sim2$.

(iv) As shown in Fig.(\ref{fig:spectraMid}) and
(\ref{fig:spectraLate}), a powerlaw component shows up in near-IR at
$t\simeq 40\spa d$ and dominates when $t\gtrsim 100\spa d$. The radio 
data \citep{2012ApJ...753...77C} is consistent with optically thin
synchrotron emission $F_\nu\propto 
\nu^{1/3}$, so the near-IR powerlaw may be due to external
shocks. However, as pointed out by  
\citet{2015ApJ...805...68P}, the sharp drop in the optical-UV lightcurves
between $181$ and $212\spa d$ (and possibly coincident with X-rays) is not
consistent with the expectations from the forward shock. A possible
explanation could be the reverse shock.
Due to possible fast cooling, the emission from the reverse
shock may track the jet kinetic power and match the observed
$t^{\sim-5/3}$ lightcurve. More radio data is needed to constrain the 
reverse shock parameters.

(v) We note the possibility that the ERF has a powerlaw instead of
blackbody spectrum as assumed in the model in this work. A powerlaw
spectrum may come from a hot corona above the disk or shocks. For
example, \citet{2012ApJ...752...18K} show that Comptonization of disk
photons by the thermal electrons at the reflected shock (due to
centrifugal barrier) adds a powerlaw extension plus Wien cut-off to
the disk SEDs at high frequencies. This mechanism alone can not
explain the X-rays in Sw J2058+05, because the temperature of the
shock-heated electrons can not reach $1-10\ keV$ (a rough estimate can
be obtained from Eq. \ref{eq:55}, if the outflowing rate $\dot{m}_w$
is replaced by accretion rate $\dot{m}$). The energy budget of the
reflected shock is also too small 
to account for the high X-ray luminosity. However, the Comptonized
powerlaw spectrum could act as the ERF for the EIC process in
the jet. If the ERF has $\nu L_\nu\propto \nu^{0.3}$, electrons in the jet
do not need to be accelerated in order to maintain a powerlaw
distribution. A self-consistent modeling of the EIC
scattering of powerlaw ERF should be done in the future.

(vi) We also note that even if the observed X-rays are from some other
processes (e.g. synchrotron emission after magnetic dissipations),
the EIC emission has typical 
luminosity of $10^{45-48}\spa erg \spa s^{-1}$ and could be
detected by the current generation of X-ray telescopes up to high
redshift $z\sim 1$. When the other processes are less efficient,
the EIC component could stand out and dominate.
Future wide field-of-view X-ray telescopes, such
as eROSITA \citep{2012arXiv1209.3114M}, Einstein
Probe\footnote{\href{http://ep-ecjm.bao.ac.cn/}{http://ep-ecjm.bao.ac.cn/}}, LOFT
\citep{2012ExA....34..415F}, will be able to find a large number of
jetted TDEs and the EIC scenario could be
tested. \citet{2015ApJ...803...36D} use Sw J1644+57 as a prototype and 
estimate the detection rates to be $0.1-10\spa yr^{-1}$ for eROSITA (up
to redshift $z_{max}\simeq  0.4$) and $1-10^2\spa yr^{-1}$  for Einstein
Probe and LOFT ($z_{max}\simeq 1$). The rates depend on the jet
beaming angle sensitively, with the upper limits coming from
$\theta_j=1/2$ ($\Gamma=2$) and the lower limit from $\theta_j = 1/20$
($\Gamma=20$). 

(vii) Lastly, we discuss the Compton drag on the jet from the EIC
process. Constraints on jet parameters can be obtained by requiring
the EIC luminosity (either from Eq. \ref{eq:6} or \ref{eq:12}) to be
smaller than the kinetic power of the jet
\begin{equation}
  \label{eq:18}
  L_{EIC} \leq L_j
\end{equation}
For simplicity, we assume $\theta_j= 1/\Gamma$ and $L_{BB} =
10^{45}\spa erg \spa s^{-1}$. The EIC
luminosity from above the photosphere $L_{EIC}^{(1)}$ (Eq. \ref{eq:6})
depends on $\tau_{j,r}(R_{ph})$, which is given by
\begin{equation}
  \label{eq:35}
  \tau_{j,r}(R_{ph}) = 0.35 \frac{L_{j,48}}{R_{ph,14.5}\Gamma_1
    \mathrm{max} (1, \sigma)}
\end{equation}
where $R_{ph,14.5} = R_{ph}/3\times10^{14}\ cm$ and $\sigma$ is the
jet magnetization. Combining
Eq.(\ref{eq:6}), (\ref{eq:18}) and (\ref{eq:35}), we obtain
\begin{equation}
  \label{eq:90}
  \Gamma\gamma_e^2 \leq \frac{2.8\times10^2}{(1-\beta_w)^2}
  R_{ph,14.5} \mathrm{max}(1, \sigma)
\end{equation}
For a typical TDE jet bulk Lorentz factor $\Gamma\sim 10$, the Compton
drag argument in Eq.(\ref{eq:90}) requires $\gamma_e\lesssim 10
(1-\beta_w)^{-1} \mathrm{max}(1, \sigma^{1/2})$ at radii $R\sim
10^{14}-10^{15}\ cm$. 

The EIC luminosity from below the photosphere $L_{EIC}^{(2)}$ is given
by Eq.(\ref{eq:12}) and we obtain from the Compton drag argument
\begin{equation}
  \label{eq:93}
  \begin{split}
      \Gamma \gamma_e^{2/3} &\leq 7.9 L_{j,48}^{1/3}
  \frac{(1-\beta_w/3)^{1/3} (1+\beta_w)}{(1-\beta_w)^{2/3}} \\
  &\cdot \mathrm{min}[1, (R_{iso}/R_{adv})^{-1/9}] \mathrm{,\ if\ }\sigma
  \lesssim 10^3
  \end{split}
\end{equation}
which depends very weakly on $\sigma$ through $R_{iso}/R_{adv}\propto
\sigma^{-1/2}$. Note that Eq.(\ref{eq:93}) is only valid when $\sigma\lesssim
10^3$, because otherwise we have
$R_{iso}\lesssim$ a few $R_{S}$ (Schwarzschild radius) and the expression of
EIC luminosity in Eq.(\ref{eq:12}) breaks down. When $\sigma\gtrsim
10^3$, the Compton drag argument can be expressed as the condition
that the EIC cooling time of individual electrons should be longer
than the dynamical time
\begin{equation}
  \label{eq:100}
  \frac{t_{EIC}}{t_{dy}} \simeq \frac{\sigma \Gamma m_p
    c^2}{\Gamma^2\gamma_e^2 U c\sigma_T} \frac{c}{R}  \geq 1
\end{equation}
where the ERF energy density can be estimated by $U\simeq L_{acc}/(4\pi
R^2 c)$ and $L_{acc}$ is the accretion luminosity
of the disk. Also, we have assumed that each electron shares a total 
energy\footnote{The momentum of a Poynting dominated jet is carried by
magnetic field (MF) comoving with baryons. The MF is ``frozen'' in the
plasma and the momentum exchange between MF and charged particles
occurs at 
the Larmor timescale (much shorter than the dynamical time). Therefore, the
bulk kinetic energy of baryons cannot drop to zero by Compton drag
on electrons, unless the momentum carried by MF, which is
coupled to charged particles, is also depleted.} of $\sigma \Gamma m_p
c^2$ and electrons' thermal Lorentz 
factor in the comoving frame is maintained at an arbitrary
$\gamma_e$. From Eq.(\ref{eq:100}), we obtain the following constraint
on jet and electron Lorentz factors
\begin{equation}
  \label{eq:38}
  \Gamma\gamma_e^2 \leq 85 \sigma_3
  \frac{ R_{12}}{L_{acc,46}} \mbox{, if } \sigma\gtrsim 10^3
\end{equation}

Any model trying to explain the X-ray data needs to take the
constraints from the Compton drag into account. For example, if the
X-rays are produced by synchrotron emission, then at least a small
fraction of jet electrons must be accelerated to Lorentz factor
$\gamma_e\gtrsim 10^3(B^\prime/10^3\ G)^{-1/2}$. The Compton drag
arguments (Eq. \ref{eq:90}, 
\ref{eq:93} and \ref{eq:38}) impose upper limits on the hot electron
fraction at the corresponding radii.


\section{Summary}
In jetted TDEs, the relativistic jet is expected to intercept a
strong external radiation field (ERF) and electrons in the jet will
inverse-Compton scatter the ERF. In this work, we calculate the
external inverse-Compton (EIC) emission from the jet.

In the case of Sw J2058+05, there is a blackbody component in
the optical-UV spectrum. We show  
that the blackbody component is consistent with being produced by a
super-Eddington wind. Using the observed blackbody component as the
ERF, we test if the X-ray luminosity and spectrum are consistent with
the EIC emission. First, to match the powerlaw spectrum $\nu L_\nu
\propto \nu^{\sim 0.3}$, electrons need to have a powerlaw distribution
$dN_e/d\gamma_e \propto 
\gamma_e^{-p}$ $(\gamma_{min} < \gamma_e < \gamma_{max})$ with $p
\simeq 2.4$. Then, we try to match the expected EIC luminosity in
the $0.3-10\ keV$ window with the observation. 
We find that for a slow wind of speed $\beta_w=v_w/c\lesssim 0.6$, the EIC
emission from above the photosphere is consistent with observations
but that from below the photosphere over-produces the X-ray
luminosity. On the other hand, if the wind is mildly relativistic with
$\Gamma_w\gtrsim1.5$, the EIC emission from both above and below the
photosphere is consistent with observations with jet parameters $\Gamma
\simeq 5-10$ and $\gamma_{min}\sim 1$.

We show that even if the observed X-rays are from some other processes 
(e.g. magnetic dissipations, see \citet{2015MNRAS.453.1820K} and
\citet{Crumley2015}), the EIC emission proposed in this work has 
typical luminosity of $10^{45}-10^{48}\spa erg \spa s^{-1}$ and could
be detected by current generation of X-ray telescopes up to high
redshift $z\sim 1$. Future wide field-of-view X-ray surveys, such as
eROSITA \citep{2012arXiv1209.3114M}, Einstein Probe, LOFT
\citep{2012ExA....34..415F} will be able to find a large number of
jetted TDEs and the EIC model could be tested.

We also show that the ERF may impose significant Compton drag on the
jet. The requirement that the Compton drag doesn't bring the jet to
a halt constrains the bulk Lorentz factor $\Gamma$ and electrons'
(thermal) Lorentz factor $\gamma_e$ in the jet comoving frame. For
example, if the jet opening angle $\theta_j = \Gamma^{-1}$ and the
thermal ERF has luminosity $10^{45}\ erg\ s^{-1}$, we find $\Gamma
\gamma_e^2\lesssim 3\times 10^2 (1-\beta_w)^{-2} \mbox{max}(1,
\sigma^{1/2})$ at $R\sim 10^{14}-10^{15}\ cm$ (the photospheric radius
of the ERF  
emitting material), where $\sigma$ is the magnetization of the
jet. Studying the EIC emission may help us to understand the   
composition of the jet and constrain the radius where the jet
energy is converted to radiation.

\section{acknowledgments}
We acknowledge helpful discussions with R.-F. Shen,
S. Markov and R. Santana. We thank the anonymous referee for a
thorough review of the paper, which helped to significantly improve
the text. This research was funded by a graduate fellowship  (``Named
Continuing Fellowship'') at the University of Texas at Austin.

\label{lastpage}
\end{document}